\documentclass[aps,prl,reprint,superscriptaddress]{revtex4-2}

\usepackage{graphicx}
\usepackage{dcolumn}
\usepackage{bm}
\usepackage[colorlinks=true,linkcolor=blue,citecolor=blue,urlcolor=blue,pdfauthor={ },pdftitle={ },pdfsubject={ },pdfkeywords={ }]{hyperref}
\usepackage{soul, color, xcolor}
\usepackage{amsmath}
\usepackage{pifont} 

\usepackage{braket}
\usepackage{lipsum}
\usepackage{scrextend}
\usepackage{upgreek}
\usepackage{mathrsfs}
\usepackage{capt-of}
\usepackage{xcolor} 
\usepackage{siunitx}
\usepackage[hints,insection]{minitoc}
\usepackage{placeins}
\usepackage{booktabs}         
\usepackage{threeparttable}    
\usepackage{multirow}    
\usepackage{diagbox} 
\usepackage{array}     
\usepackage{ragged2e}
\usepackage{xspace}

 \iftrue

\else

\fi
\newcommand{\smtext}{Supplemental Material} 
\newcommand{\smref}[1]{see \smtext~#1} 
\providecommand{\DIFdel}[1]{}

\newcommand{\addressa}{\affiliation{Laboratory of Quantum Information, University of Science and Technology of China, Hefei, Anhui, 230026, China}}
\newcommand{\addressb}{\affiliation{CAS Center For Excellence in Quantum Information and Quantum Physics, University of Science and Technology of China, Hefei, Anhui, 230026, China}}
\newcommand{\addressc}{\affiliation{Institute of Artificial Intelligence, Hefei Comprehensive National Science Center, Hefei, Anhui, 230088, China}}

\newcommand{\addresse}{\affiliation{Origin Quantum, Hefei, Anhui 230088, China}}
\newcommand{\addressf}{\affiliation{Suzhou Institute for Advanced Research, University of Science and Technology of China, Suzhou, Jiangsu 215123, China}}

\begin{document}

\title{Spectator Leakage Suppression via Invariant Subspace Engineering for CZ Gates in Superconducting Quantum Circuits}

\author{Peng Wang}
\addressa\addressb\addressf
\author{Bin-Han Lu}
\author{Tian-Le Wang}
\addressa\addressb
\author{Sheng Zhang}
\addressa\addressb\addressf
\author{Zhao-Yun Chen}
\addressc
\author{Hai-Feng Zhang}
\author{Ren-Ze Zhao}
\author{Xiao-Yan Yang}
\author{Ze-An Zhao}
\addressa\addressb
\author{Zhuo-Zhi Zhang}
\author{Xiang-Xiang Song}
\addressa\addressb\addressf
\author{Yu-Chun Wu}
\addressa\addressb\addressc
\author{Peng Duan}
\email{pengduan@ustc.edu.cn}
\addressa\addressb

\author{Guo-Ping Guo}
\email{gpguo@ustc.edu.cn}
\addressa\addressb\addresse

\begin{abstract}

Spectator leakage poses a fundamental challenge to scalable quantum computing, particularly as frequency collisions become unavoidable in multi-qubit processors. We introduce a leakage mitigation strategy based on dynamically reshaping the system Hamiltonian. Our technique utilizes a tunable coupler to enforce a block-diagonal structure on the effective Hamiltonian governing near-resonant spectator interactions, confining the gate dynamics to a two-dimensional invariant subspace and thus preventing leakage by construction. 
On a multi-qubit superconducting processor, we experimentally demonstrate that this dynamic control scheme suppresses leakage rates to the order of $10^{-4}$, across a wide near-resonant detuning range and with up to three simultaneous spectator qubits. 
These results demonstrate a robust and scalable method that resolves the critical trade-off between dense frequency packing and high-fidelity gate operation. Our work establishes dynamic Hamiltonian engineering as an essential technology for building fault-tolerant quantum computers.
\end{abstract}

\maketitle

\textit{Introduction.—} 
Scalable quantum information processing requires high-fidelity operations within increasingly complex many-body systems~\cite{2018_preskillquantum,2020_kjaergaardsuperconducting,2023_atomhighfidelityparallel}. As systems scale and gate parallelism increases, frequency crowding emerges as a fundamental physical obstacle~\cite{2021_freq_crowd,2022_zhanghigh,2016_theissimultaneous}. The inevitable rise in the density of energy levels leads to frequency collisions and unwanted near-resonant interactions~\cite{2018_brinkdevice,2023_crosstalk_xy, 2024_crosstalk_xy, 2018_coupler_zz, 2022_coupler_zz, 2022_quamtum_crosstalk_sc}. These coherent error channels are challenges common to many quantum architectures~\cite{2025_cross_review, 2021_crosstalk_ion, 2022_crosstalk_ion, 2019_parallel_atom,2023_surface_code_google,2023_leakage_spread}. While significant progress has been made in mitigating incoherent noise, managing the coherent dynamics in such spectrally congested environments remains a critical bottleneck for large-scale architectures~\cite{2025_blandMillisecondLifetimesCoherence,2024_ParkPassive,2021_WangSingle,2024_ferracinefficiently}.

Among these coherent errors, leakage to non-computational spectator states is particularly detrimental~\cite{2022_quamtum_crosstalk_sc, 2023_surface_code_google}. In the context of superconducting circuits, these errors manifest in controlled-Z (CZ) gates when the spectator qubit enters a near-resonant condition with one of the gate qubits, inducing coherent population leakage via transverse exchange coupling~\cite{2020_spectator_cz, 2021_spectator_cr,2021_spectator_zajac, 2026_spectator_cz}. Historically, in small- to medium-scale systems, frequency allocation algorithms have typically treated near-resonance as a hard constraint, effectively avoiding such conditions through careful qubit frequency planning~\cite{2020_snake, 2024_freq_allo_google, 2024_freq_allo_lbh, 2025_freq_allo}. While active mitigation techniques exist, they often require large detuning or specific parameter regimes~\cite{2020_spectator_cz,2022_QECd3}. However, with the continued scaling of quantum processors and the increasing demand for high-connectivity layouts tailored to efficient quantum error correction codes~\cite{2026_bbcode_zju, 2024_bbcode_ibm}, frequency collisions are becoming increasingly difficult to avoid. As a result, near-resonant spectator conditions are likely to occur with higher probability, posing a significant threat to overall system performance. A systematic understanding of this error mechanism, along with the development of robust mitigation strategies, is therefore of pressing importance.

To address spectator leakage, we develop a scalable mitigation strategy based on invariant subspace engineering. Specifically, we reshape the effective interaction during near-resonant spectator conditions, enforcing a block-diagonal Hamiltonian that confines gate dynamics to a closed two-dimensional subspace and intrinsically prevents leakage.
We construct a general three-level model to capture the relevant leakage channel and verify the protocol via numerical simulations. Using a superconducting processor featuring flux-tunable transmon qubits and tunable couplers as a versatile experimental platform, we demonstrate leakage suppression down to the $10^{-4}$ level. The method remains effective in the presence of multiple simultaneous spectators, highlighting its scalability and practical relevance to fault-tolerant quantum computing.

\textit{Leakage-Free Subspace Engineering.—} 
The dynamic non-adiabatic CZ gate brings the states $\ket{11}$ and $\ket{02}$ into resonance, inducing a two-level Rabi oscillation that drives $\ket{11}$ through a round-trip evolution and imparts a conditional $\pi$ phase~\cite{2003_cz_theory, 2009_cz_exp}. However, when a spectator qubit becomes near-resonant with one of the gate-involved transitions, the effective dynamics extend beyond this simple two-level picture.

\begin{figure}[htbp]
    \centering
    \includegraphics[width=\linewidth]{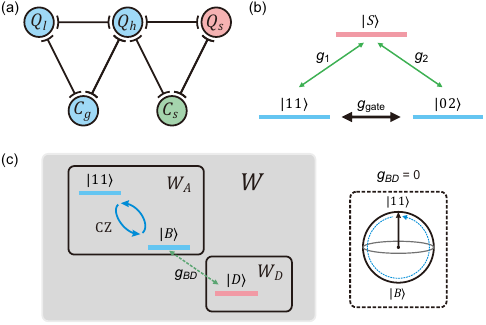}
    \caption{(a) Schematic representation of the three-qubit architecture, comprising two gate qubits $Q_l$, $Q_h$ and a spectator qubit $Q_s$ interconnected via nearest-neighbor coupling and mediated by tunable couplers $C_g$ and $C_s$. (b) The effective three-level model describing the near-resonant interaction between the states $\ket{11}$, $\ket{02}$, and the spectator state $\ket{S}$, with couplings $g_{\mathrm{gate}}$, $g_1$, and $g_2$. (c) Conceptual illustration of the block-diagonalization strategy. The original three-level space is decomposed into two invariant subspaces, $W_A$ and $W_D$. By dynamically tuning the system to enforce $g_{BD}=0$, the evolution is confined to the subspace $W_A = \{\ket{11}, \ket{B}\}$, effectively restoring the two-level CZ gate dynamics governed by a Rabi oscillation between $\ket{11}$ and $\ket{B}$, as depicted on the Bloch sphere.
    }
    \label{fig:theory_schematic}
\end{figure}

Consider a system of a lower-frequency qubit $Q_l$, a higher-frequency qubit $Q_h$ and a spectator qubit $Q_s$ interconnected via tunable couplers ($C_g$, $C_s$) as schematically shown in Fig.~\ref{fig:theory_schematic}(a). We define the state order as $|lhs\rangle$($\ket{lh}$) to avoid confusion. By applying a Schrieffer-Wolff (SW) transformation, the dynamics under a specific near-resonant condition can be truncated to a three-dimensional subspace comprising the states $\ket{11}$, $\ket{02}$, and a spectator state $\ket{S}$. The interactions within this subspace are depicted in Fig.~\ref{fig:theory_schematic}(b), and the Hamiltonian takes the general form:
\begin{align}
    H_{\mathrm{eff}} =
    \begin{pmatrix}
    \omega_{11} & g_{\mathrm{gate}} & g_1 \\
    g_{\mathrm{gate}} & \omega_{02} & g_2 \\
    g_1 & g_2 & \omega_{S}
    \end{pmatrix}.
    \label{eq:H_eff_general}
\end{align}
Here, $\omega_{11}$, $\omega_{02}$, and $\omega_{S}$ are the energies of the respective states with $\omega_{11} = \omega_{02}$. $g_{\mathrm{gate}}$ is the primary gate coupling between $\ket{11}$ and $\ket{02}$, while $g_1$ and $g_2$ are the undesirable couplings to the spectator state $\ket{S}$. The presence of non-zero $g_1$ and $g_2$ couples the three states, making leakage from $\ket{11}$ to $\ket{S}$ and $\ket{02}$ possible. The physical origin of $\ket{S}$, and the values of $g_1$ and $g_2$, depend on the specific spectator state and system architecture. While implemented here on a superconducting architecture, this effective model describes a generic class of three-level dynamics ubiquitous in quantum platforms.

A common approach to manage such multi-level dynamics is to find specific gate durations where the desired conditional phase is acquired while leakage is simultaneously suppressed through temporal interference~\cite{barendsDiabaticGatesFrequencyTunable2019}. Synchronizing these two conditions, however, does not necessarily yield the shortest possible gate time. We propose an alternative and more direct strategy: dynamically reconfiguring the Hamiltonian to enforce a block-diagonal structure, as illustrated in Fig.~\ref{fig:theory_schematic}(c). Our goal is to find a two-dimensional invariant subspace $W_A$ that contains the initial state $\ket{11}$ and remains closed under the evolution governed by $H_{\mathrm{eff}}$. This confines the gate dynamics and intrinsically prevents leakage to its orthogonal complement.

To achieve this, we seek a basis transformation that decouples one of the new basis vectors from the other two. As we require the computational state $\ket{11}$ to be part of the desired two-dimensional invariant subspace, the transformation should only mix the other two states, $\ket{02}$ and $\ket{S}$. This is achieved with a rotation in the $\{\ket{02}, \ket{S}\}$ subspace, defining a ``bright'' state $\ket{B}$ and a ``dark'' state $\ket{D}$:
\begin{align}
\ket{B} &= \cos{\theta}\ket{02} + \sin{\theta}\ket{S}, \label{eq:B_state_new} \\
\ket{D} &= -\sin{\theta}\ket{02} + \cos{\theta}\ket{S}. \label{eq:D_state_new}
\end{align}
In this new basis $\{\ket{11}, \ket{B}, \ket{D}\}$, the transformed Hamiltonian $H'_{\mathrm{eff}}$ will be block-diagonal if the coupling between the $\{\ket{11}, \ket{B}\}$ subspace and the state $\ket{D}$ is zero. This requires two conditions to be met: $\bra{D} H'_{\mathrm{eff}} \ket{11} = 0$ and $\bra{D} H'_{\mathrm{eff}} \ket{B} = 0$.

The first condition, decoupling $\ket{11}$ from $\ket{D}$, is satisfied by choosing the rotation angle $\theta = \tan^{-1} \frac{g_1}{g_{\mathrm{gate}}}$. With this specific angle, the second condition for block-diagonalization, $\bra{D} H'_{\mathrm{eff}} \ket{B} = 0$, imposes a constraint on the system parameters:
\begin{align}
    g_{BD}=\frac{\omega_S - \omega_{11}}{2}\sin{2\theta} + g_2\cos{2\theta} = 0.
    \label{eq:g2_condition}
\end{align}

Once this condition is met, the Hamiltonian is block-diagonalized. Eq.~\eqref{eq:g2_condition} reveals that leakage cancellation arises from the destructive interference between detuning and coupling terms. Consequently, this block-diagonalization strategy is broadly applicable to any controlled quantum system where the effective interaction strength or energy detuning can be dynamically tuned. The dynamics relevant to the CZ gate are then confined to the two-dimensional invariant subspace $W_A = \{\ket{11}, \ket{B}\}$, governed by a simple $2 \times 2$ Hamiltonian:
\begin{align}
	H_{A} =
	\begin{pmatrix}
		\omega_{11} & g_{B} \\
		g_{B} & \omega_{B}
	\end{pmatrix},
	\label{eq:H_A}
\end{align}
where $g_B = \sqrt{g_{\mathrm{gate}}^2 + g_1^2}$ is the new effective gate coupling and $\omega_{B}$ is the frequency of the bright state. The evolution becomes that of a standard CZ gate, albeit with a slightly modified coupling strength and interaction partner, and leakage is exactly nulled when the gate duration $\tau = 2\pi/\sqrt{g_B^2 + \left(\omega_{11} - \omega_B \right)^2}$ by construction. 
After one complete cyclic evolution in the ${|11\rangle,|B\rangle}$ subspace, the corresponding effective unitary reduces to the standard CZ matrix in the computational subspace , as shown explicitly in Supplemental Material I.C~\cite{supp}. More generally, the same closed-subspace picture can be extended to detuned cyclic evolution, providing a route to arbitrary controlled-phase gates.

While this condition can only be met accidentally in systems with fixed couplings, our architecture with a tunable spectator coupler provides the necessary control of $g_\text{BD}$ by shifting the spectator coupler frequency to an optimal ``off'' point during the CZ gate. 
\begin{figure}[htbp]
	\centering
	\includegraphics[width=\linewidth]{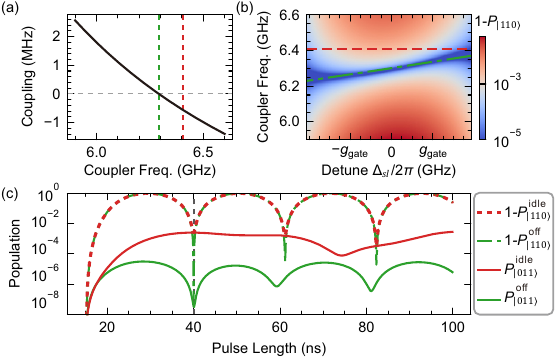}
	\caption{Numerical simulation of the leakage suppression scheme. The simulation models a 40~ns CZ gate operating at $\omega_{l}^{CZ}/2\pi=4.276$~GHz. The relevant couplings are the primary gate coupling $g_{\mathrm{gate}}/2\pi=27$~MHz, a fixed leakage coupling $g_1/2\pi \approx 0.5$~MHz, and a tunable leakage coupling $g_2$ controlled via the spectator coupler frequency.
		(a) The calculated leakage coupling $g_{BD}$ as a function of the spectator coupler frequency. The black solid line shows that $g_{BD}$ can be continuously tuned and precisely nulled at a specific frequency. The vertical dashed lines correspond to the coupler frequencies used for the time-evolution simulations in (c).
		(b) Final leakage population at the end of a CZ gate. The green dashed line marks the condition $g_{BD}=0$ calculated from (a). The horizontal red dashed line indicates the coupler's idle frequency.
		(c) Final population as a function of gate length for two different coupler settings: idle (non-optimal) and off (optimal). At the ``off'' frequency, this leakage population undergoes coherent oscillations characteristic of a closed two-level system.
	}
	\label{fig:simulation_results}
\end{figure}

To validate our theoretical framework and the effectiveness of our approach, we simulate the three-level Hamiltonian in Eq.~\eqref{eq:H_eff_general} under a realistic l--h--s topology, where the spectator qubit $Q_s$ is near-resonant with the gate qubit $Q_l$. When the spectator is initialized in its ground state, leakage predominantly arises from the coupling between the gate states $\ket{110}$, $\ket{020}$ and the spectator state $\ket{011}$. As shown in Fig.~\ref{fig:simulation_results}(a), $g_\text{BD}$ varies with the spectator coupler frequency, and the ``off'' point where $g_\text{BD} = 0$ can be directly identified. At this point, the gate evolution is confined to an effective two-level subspace, and leakage is suppressed by construction. This is clearly evidenced in Fig.~\ref{fig:simulation_results}(b), where a distinct valley of minimal leakage aligns exactly with the predicted condition. In Fig.~\ref{fig:simulation_results}(c), time-domain simulations further reveal that, under the idle condition, the population in the spectator state $\ket{011}$ exhibits high-amplitude, slow oscillations, indicating complex multi-level dynamics. In contrast, the off condition yields a clean monochromatic oscillation, consistent with an effective two-level model. At the end of the gate cycle (40~ns), the total leakage under the idle condition is dominated by population in the spectator state $\ket{011}$, whereas both are simultaneously and significantly suppressed at the off frequency.

This framework can be extended to scenarios with multiple spectator levels, where the goal becomes finding a two-dimensional invariant subspace within a larger Hilbert space, a task achievable if sufficient control over the relevant couplings is available (\smref{I.D}~\cite{supp}).

\textit{Experiments.—}
We implemented the protocol on the superconducting processor \textit{Wukong}~\cite{2025_Qrouter}.
The device features a flip-chip architecture with flux-tunable transmon qubits.
Nearest-neighbor qubits are interconnected via tunable couplers, which are also designed as flux-tunable asymmetric transmons.
The effective interaction between qubits is controlled by modulating the magnetic flux threading the coupler's SQUID loop, which adjusts the coupler frequency relative to the qubits~\cite{2018_coupler_zz}.
To mitigate idle point collisions, we implemented a frequency allocation scheme that strategically separates operating frequencies of the gate and spectator qubits during parallel operations~\cite{2025_allo_nn}.

Our investigation proceeds as follows. We first established a baseline by calibrating a 40~ns CZ gate, implemented with a flat-top Gaussian pulse, while keeping the spectator qubit far detuned. To study the spectator leakage, we then dynamically tuned the spectator into the resonance region of the gate qubits and applied our dynamic coupler technique to mitigate the leakage. Finally, we evaluated the scalability of this approach in scenarios involving multiple simultaneous spectator qubits.

\begin{figure}[htbp]
    \centering
    \includegraphics[width=\linewidth]{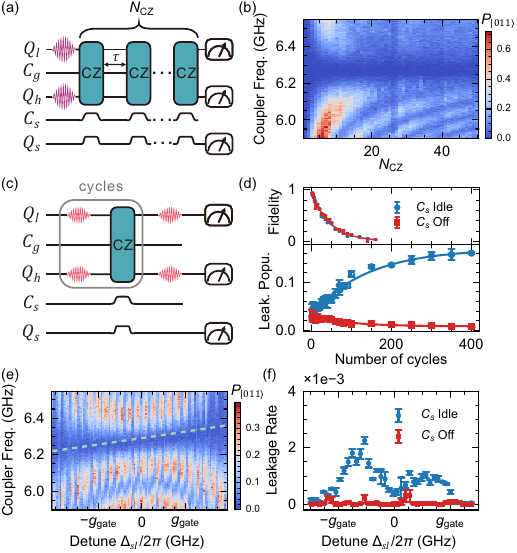}
    \caption{Experimental validation of the dynamic leakage suppression protocol.
    (a) Pulse sequence for amplifying leakage. A delay $\tau$ between consecutive CZ gates enables constructive interference. 
    (b) Measured leakage population $P_{011}$ as a function of the spectator coupler frequency and the number of CZ $N_{CZ}$, used to identify the optimal ``off'' frequency that minimizes leakage. 
    (c) Schematic of the XEB sequence used for characterizing gate fidelity and leakage. During the CZ gate operation, the spectator qubit is tuned to resonance at 4.27 GHz.
    (d) XEB fidelity and leakage population are measured for the coupler frequency set to the off and idle points. Solid lines are fits to the data, used to extract gate errors and leakage rates.The observed decay in leakage population for the optimized configuration arises from the equilibration process where the initial thermal population exceeds the steady-state limit. Detailed analysis confirms this behavior is consistent with the spectator relaxation dynamics (see Supplemental Material III.C~\cite{supp}).
    (e) Optimal coupler off frequencies measured across a range of spectator detunings. The stripes in the plot arise from sampling rate limitations. The experimental data (blue valley) shows good agreement with the theoretical prediction (Light green dashed line) from our model.
    (f) Leakage rates are measured for the spectator coupler at the off and idle frequency across various spectator qubit frequencies $\omega_s$.
    }
    \label{fig3}
\end{figure}
To accurately characterize the off-frequency of the coupler, we measured the leakage in the spectator qubit as the gate operation progressed. 
Since leakage in a single CZ gate is small, we utilized the pulse sequence in Fig.~\ref{fig3}(a), which introduces a controlled delay between successive CZ gates, amplifying the leakage via constructive interference~\cite{2020_floquet, 2024_floquet}. By preparing the system in the $\ket{110}$ state and measuring the population transferred to the $\ket{011}$ state, we scanned the coupler frequency to find the point of minimum leakage, as shown in Fig.~\ref{fig3}(b).

We then used the cross-entropy benchmarking (XEB) sequence to validate the effectiveness of leakage suppression~\cite{2018_xeb_quan_supremacy}. The spectator qubit was dynamically tuned into resonance only during the CZ operations as shown in Fig.~\ref{fig3}(c). 
Following the sequence, we measured the cross-entropy in the two-qubit computational space and leakage population in the spectator state to assess gate fidelity and leakage rate simultaneously. The results are presented in Fig.~\ref{fig3}(d).
From fits to the leakage population data (bottom panel), we extracted the per-gate leakage rate $L_1$. The rate was suppressed by a factor of approximately 30, from $1.21(5)\times10^{-3}$ down to $4(3)\times10^{-5}$. This final leakage rate of $\sim 10^{-5}$ approaches the resolution limit of our setup, which is constrained by readout fidelity and spectator decoherence.
Concurrently, from fits to the XEB fidelity data (top panel), we extracted the total CZ gate error $\epsilon$. The error was reduced from $1.52\left(3 \right)\times10^{-2}$ to $1.34\left(10 \right)\times10^{-2}$. This final error is consistent with the gate's baseline performance in the dispersive regime, where the spectator is far detuned. Notably, the magnitude of this error reduction is consistent with the independently measured contribution from the suppressed leakage rate $L_1$ (\smref{III.D}~\cite{supp}). This confirms that our method specifically targets and eliminates the leakage-induced error pathway, without introducing other significant detrimental effects.

To assess the robustness of our method, we systematically characterized the optimal coupler off frequencies across a range of spectator frequencies. As shown in Fig.~\ref{fig3}(e), for any given $\omega_s$ in the near-resonant region, we can experimentally identify a corresponding off frequency, and the trend shows agreement with the predictions of our theoretical model. We then confirmed the effectiveness of this dynamic tuning in Fig.~\ref{fig3}(f), showing that leakage is uniformly suppressed to the order of $10^{-4}$ across the near-resonant detuning range. This capability effectively eliminates the need for spectator qubits to remain in the far-detuned dispersive regime, significantly alleviating frequency collision constraints in scalable architectures. The consistent, wide-range suppression highlights the robustness and practical applicability of our dynamic control scheme.

\begin{figure}[htbp]
    \centering
    \includegraphics[width=\linewidth]{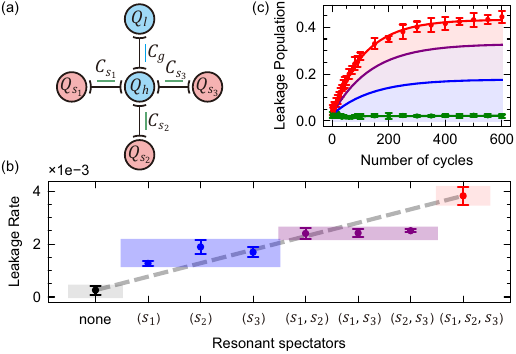}
    \caption{(a) Schematic diagram of the 5 qubits system used to study leakage mitigation with multiple spectator qubits. (b) Measured leakage rates of the CZ gate for different configurations of spectator qubits. The horizontal axis represents the combinations of spectator qubits tuned to the resonance region, with the others at the idle point. (c) Leakage population in the non-computational subspace as a function of the number of XEB cycles. The red and green lines represent the cases where all three spectator qubits are tuned to resonance, with the spectator coupler set to the idle and off points, respectively. The colored fillings indicate the leakage contributions from each spectator qubit, with blue, purple, and red representing the contributions from $Q_{s_{1}}$, $Q_{s_{2}}$, and $Q_{s_{3}}$, respectively.
    }
    \label{fig4}
\end{figure}

To demonstrate the scalability of our method, we investigated its performance in the presence of multiple simultaneous spectator qubits, as depicted in Fig.~\ref{fig4}(a). The core assumption is that the leakage contribution from each spectator is independent and linearly additive, as interactions between spectators are higher-order effects and can be safely ignored.
First, we characterized the unmitigated leakage by systematically tuning different combinations of the three spectator qubits ($Q_{s_{1}}, Q_{s_{2}}, Q_{s_{3}}$) into resonance. As shown in Fig.~\ref{fig4}(b), the total leakage rate exhibits a clear stepwise increase with the number of resonant spectators. This linear additivity, further confirmed by component analysis in the \smtext~III.G, validates our independence assumption and highlights the increasing threat of leakage in larger systems.
Next, we tested our suppression protocol in the most challenging scenario, with all three spectators simultaneously tuned to their resonant frequencies. For each spectator, we applied its independently characterized optimal coupler pulse. The results are shown in Fig.~\ref{fig4}(c). Without suppression, the total leakage rate reached $3.8\left(3\right)\times 10^{-3}$. With our dynamic control, the total leakage was suppressed to $1.9\left(4\right)\times 10^{-4}$. This demonstrates that even with multiple spectators, the total leakage remains well suppressed--reaching levels comparable to those of state-of-the-art isolated CZ gates~\cite{2021_CZ_iswap_zzfree, 2024_2q_2tran}.
While this suppressed leakage rate is higher than the single-spectator case, this is primarily attributed to the fidelity limitations of multi-qubit state readout, which makes resolving extremely low total leakage across multiple channels challenging. 
To further validate the robustness and universality of our protocol, we performed statistical characterization on two additional 5-qubit regions selected from a 72-qubit processor.
As detailed in Supplemental Material III.H~\cite{supp}, the aggregated data from 21 distinct spectator configurations confirm that the leakage suppression performance is consistent and reproducible across different device parameters.
These findings confirm that our method is highly scalable, effectively preventing the proliferation of spectator-induced leakage as system size increases.

\textit{Conclusions.—}
In summary, we have developed and experimentally demonstrated a dynamic-Hamiltonian-engineering strategy to mitigate leakage induced by spectral crowding. By constructing a two-dimensional invariant subspace, our technique block-diagonalizes the near-resonant three-level Hamiltonian, intrinsically preventing leakage for the targeted pathway. We demonstrate leakage suppression to the order of $10^{-4}$ and effective scalability with multiple spectators. This approach provides a robust solution to a critical class of coherent errors, easing frequency crowding constraints and advancing the development of fault-tolerant quantum computers. 

Our work also highlights the complex, state-dependent nature of spectator leakage. While this work focused on a leakage channel originating from a spectator in its ground state, our theoretical framework is general, as shown in the Supplemental Material I.F~\cite{supp}. Extending this block-diagonalization concept to even more complex multi-level interactions is another promising direction. Importantly, the physical principle underlying our protocol is generic and applicable to other quantum many-body systems facing similar frequency crowding. Consequently, this strategy can be adapted to these systems by tuning relevant controls, such as barrier gate voltages in semiconductor quantum dots or optical Stark shifts in atomic systems.
This work establishes invariant subspace engineering as a versatile tool for high-fidelity control in dense spectral environments, essential for the continued scaling of quantum information processing.

\begin{acknowledgments}
\textit{Acknowledgments.—}
We thank Prof. Chang-Ling Zou for reviewing the manuscript and providing valuable suggestions. 
This work has been supported by the National Key Research and Development Program of China (Grant No.~2023YFB4502500) and the National Natural Science Foundation of China (Grant No.~12034018).
This work is partially carried out at the USTC Center for Micro and Nanoscale Research and Fabrication. 
\end{acknowledgments}

%

\end{document}


\title{Supplementary Materials for ``Spectator Leakage Suppression via Invariant Subspace Engineering for CZ Gates in Superconducting Quantum Circuits''}

\author{Peng Wang}
\addressa\addressb\addressf
\author{Bin-Han Lu}
\author{Tian-Le Wang}
\addressa\addressb
\author{Sheng Zhang}
\addressa\addressb\addressf
\author{Zhao-Yun Chen}
\addressc
\author{Hai-Feng Zhang}
\author{Ren-Ze Zhao}
\author{Xiao-Yan Yang}
\author{Ze-An Zhao}
\addressa\addressb
\author{Zhuo-Zhi Zhang}
\author{Xiang-Xiang Song}
\addressa\addressb\addressf
\author{Yu-Chun Wu}
\addressa\addressb\addressc
\author{Peng Duan}
\email{pengduan@ustc.edu.cn}
\addressa\addressb

\author{Guo-Ping Guo}
\email{gpguo@ustc.edu.cn}
\addressa\addressb\addresse

\maketitle
\tableofcontents

\section{Theoretical Framework for Leakage Suppression}
\subsection{Effective Hamiltonian via Schrieffer-Wolff Transformation}
To provide a theoretical foundation for the effective three-level model used in the main text, we start with the full Hamiltonian of the multi-qubit system. We consider a system composed of gate qubits ($Q_l, Q_h$), a spectator qubit ($Q_s$), and their mediating couplers ($C_g, C_s$). The full system Hamiltonian can be written as:
\begin{align}
H=&\sum_i \left( \omega_i a^{\dag}_i a_i + \frac{\alpha_i}{2} a^{\dag}_i a^{\dag}_i a_i a_i \right)\\
&- \sum_{j=l, h } g_{j c_g}\left(a_j^\dag -a_j \right) \left(a_{c_g}^\dag - a_{c_g} \right) - g_{l h}\left(a_l^\dag -a_l \right) \left(a_h^\dag - a_h \right)\\
&- \sum_{k=h, s } g_{k c_s}\left(a_k^\dag -a_k \right) \left(a_{c_s}^\dag - a_{c_s} \right) - g_{h s}\left(a_h^\dag -a_h \right) \left(a_s^\dag - a_s \right)\\
&- g_{l s}\left(a_l^\dag -a_l \right) \left(a_s^\dag - a_s \right)
\end{align}
where $i\in \{ l, h, s, c_g, c_s \}$. Here, $\omega_i$ and $\alpha_i$ are the bare frequency and anharmonicity of each mode, respectively, $g_{jk}$ is the direct coupling strength, and $a_i (a_i^\dag)$ is the corresponding annihilation (creation) operator. Specifically, $g_{ls}$ is the next-nearest-neighbor stray coupling between $Q_l$ and $Q_s$, which typically has a magnitude on the order of 0.5~MHz in superconducting quantum circuits~\cite{2024_PST_zju}.

To derive the effective interactions in the qubit subspace, we employ the Schrieffer-Wolff (SW) transformation~\cite{2011_SW}. This perturbative method eliminates high-energy degrees of freedom, such as off-resonant couplers, yielding an effective Hamiltonian for the low-energy qubit subspace, $H_{\mathrm{eff}} = e^S H e^{-S}$. For a Hamiltonian $H=H_0+V$, where $H_0$ is diagonal and $V$ is the perturbation, $S$ is chosen to satisfy $\left[H_0, S \right]=-V$ to first order. The effective Hamiltonian up to second order is then $H_\mathrm{eff}=H_0 + \frac{1}{2} \left[S, V \right]$. 

First, we consider the interaction between the two gate qubits, $Q_l$ and $Q_h$, mediated by the  gate coupler $C_g$. The interaction term is $V_g= \sum_{j\neq k} g_{j k}\left(a_j -a_j^\dag \right) \left(a_k + a_k^\dag \right) $ where $j,k \in \{ l, h, c_g \}$. After applying the SW transformation and projecting onto the subspace where the coupler $C_g$ remains in its ground state, we obtain various effective qubit-qubit interactions within the two-excitation subspace of interest
\begin{align}
H_\mathrm{int}^g = g_{ls}^{20} \left( \ket{11} \bra{20} + \ket{20} \bra{11} \right) + g_{ls}^{02} \left( \ket{11} \bra{02}+ \ket{02} \bra{11} \right) ,
\label{eq:g_gate_appendix}
\end{align}
where the coupling between different transitions is
\begin{align}
    g_{lh}^{20} &= \sqrt{2}g_{lh} +\frac{g_{lc_g}g_{hc_g}}{\sqrt{2}}\left(\frac{1}{\Delta_{lc_g}-\alpha_l} +\frac{1}{\Delta_{hc_g}}- \frac{1}{\Sigma_{lc_g}-\alpha_l}-\frac{1}{\Sigma_{hc_g}}\right),\\
    g_{lh}^{02} &= \sqrt{2}g_{lh} +\frac{g_{lc_g}g_{hc_g}}{\sqrt{2}}\left(\frac{1}{\Delta_{lc_g}} +\frac{1}{\Delta_{hc_g}-\alpha_h}- \frac{1}{\Sigma_{lc_g}}-\frac{1}{\Sigma_{hc_g}-\alpha_h}\right).
\end{align}
where $\Delta_{ij} = \omega_i - \omega_j$ and $\Sigma_{ij} = \omega_i + \omega_j$. 

Similarly, we consider the interactions involving the spectator qubit $Q_s$, which is coupled to $Q_h$ via the spectator coupler $C_s$, that is, the l-h-s topology, discussed in the main text. The effective interaction Hamiltonian can be derived using a similar SW transformation
\begin{align}
H_\mathrm{int}^s = &g_{hs}^{01} \left(\ket{110} \bra{101} + \ket{101} \bra{110} \right) + g_{hs}^{20} \left( \ket{011} \bra{020} + \ket{020} \bra{011} \right) \notag\\
&+ g_{hs}^{02} \left( \ket{011} \bra{002}+ \ket{002} \bra{011} \right)  + g_{ls}^{01} \left(\ket{110} \bra{011} + \ket{011} \bra{110} \right),
\label{eq:H_s_appendix}
\end{align}
where the coupling between different transitions is
\begin{align}
    g_{hs}^{01} &= g_{hs} +\frac{g_{hc_s}g_{sc_s}}{2}\sum_{j=h,s}\left(\frac{1}{\Delta_{jc_s}} - \frac{1}{\Sigma_{jc_s}}\right),\\
    g_{hs}^{20} &= \sqrt{2}g_{hs} +\frac{g_{hc_s}g_{sc_s}}{\sqrt{2}}\left(\frac{1}{\Delta_{hc_s}-\alpha_h} +\frac{1}{\Delta_{sc_s}}- \frac{1}{\Sigma_{hc_s}-\alpha_h}-\frac{1}{\Sigma_{sc_s}}\right),\label{eq:g_s_appendix}\\
    g_{hs}^{02} &= \sqrt{2}g_{hs} +\frac{g_{hc_s}g_{sc_s}}{\sqrt{2}}\left(\frac{1}{\Delta_{hc_s}} +\frac{1}{\Delta_{sc_s}-\alpha_s}- \frac{1}{\Sigma_{hc_s}}-\frac{1}{\Sigma_{sc_s}-\alpha_s}\right).
\end{align}
The effective coupling between $Q_l$ and $Q_s$, $g_{ls}^{01}$ is more complex, as it receives contributions from multiple pathways. It includes not only the direct coupling $g_{ls}$ but also higher-order, coupler-mediated pathways. A significant mediated contribution comes from a fourth-order process~\cite{2024_flo_perturbation}. Combining these terms, the total effective coupling between $\ket{110}$ and $\ket{011}$ is:
\begin{align}
    g_{ls}^{01} &= g_{ls} - \frac{g_{\mathrm {gate}}g_{sc_s} g_{hc_s}}{\Delta_{hc_s} \Delta_{lc_s}}.
    \label{eq:g1_derivation}
\end{align}

We emphasize that even without direct next-nearest-neighbor coupling $g_{ls}$, this coupler-mediated interaction can be significant when $g_{\mathrm{gate}}$ is large. Its presence implies that within the three-level model, simply nulling a specific lower-order coupling such as an effective $g_{hs}^{02}$ may not suffice to fully decouple the spectator level, as this $g_{ls}^{01}$ pathway in $\ket{11}$ and $\ket{S}$. 

Combining the relevant interaction terms, we can now construct the specific Hamiltonian for the l-h-s topology with the spectator state $\ket{S} = \ket{011}$. In the basis $\{\ket{110}, \ket{020}, \ket{011}\}$, the effective Hamiltonian is
\begin{align}
    H_{\mathrm{eff}} = 
    \begin{pmatrix}
    \omega_{110} & g_{lh}^{02} & g_{ls}^{01} \\
    g_{lh}^{02} & \omega_{020} & g_{hs}^{20} \\
    g_{ls}^{01} & g_{hs}^{20} & \omega_{011}
    \end{pmatrix}.
\end{align}
This matrix is the explicit form of the general Hamiltonian presented in Eq.~(1) of the main text. The parameters are therefore identified as
\begin{align}
    g_{\mathrm{gate}} = g_{lh}^{02}, g_1 = g_{ls}^{01}, g_2 = g_{hs}^{20}.
\end{align}
Crucially, the coupling $g_{hs}^{20}$ is tunable via the spectator coupler frequency $\omega_{c_s}$ as shown in Eq.~\eqref{eq:g_s_appendix}, providing the necessary control for the leakage suppression protocol.

\subsection{Bright and Dark State Basis Transformation}
To analyze the leakage pathway described by the effective Hamiltonian $H_{\mathrm{eff}}$, we transform it into the bright and dark state basis. This basis is chosen to isolate the dynamics relevant to the CZ gate. The transformation operates on the $\{\ket{02}, \ket{S}\}$ subspace and can be described by the unitary operator
\begin{align}
    U^{-1} = \begin{pmatrix}
    1 & 0 & 0 \\
    0 & \cos\theta & \sin\theta \\
    0 & -\sin\theta & \cos\theta
    \end{pmatrix},
\end{align}
where the mixing angle $\theta = \arctan(g_1/g_{\mathrm{gate}})$ is chosen to ensure that the state $\ket{11}$ only couples  couples to $\ket{B}$.

The transformed Hamiltonian $H'_{\mathrm{eff}} = U^\dag H_{\mathrm{eff}} U$ in the new basis $\{\ket{11}, \ket{B}, \ket{D}\}$ takes the form:
\begin{align}
\label{eq:H_tilde}
H'_{\mathrm{eff}} &= {%
    \begin{pmatrix}
    \omega_{11} & g_B & 0 \\
    g_B & \omega_{B} & g_{BD} \\
    0 & g_{BD} & \omega_{D}
    \end{pmatrix}
},
\end{align}
The energies of the bright and dark states, $\omega_B$ and $\omega_D$, are
\begin{align}
\omega_B &= \omega_{11} + \Delta_{sl}\sin^2\theta + g_{2}\sin{2\theta}, \label{eq:omega_B}\\
\omega_D &= \omega_{11} + \Delta_{sl}\cos^2\theta - g_{2}\sin{2\theta}. \label{eq:omega_D}
\end{align}
The coupling strengths are 
\begin{align}
g_B &= \sqrt{g_{\mathrm{gate}}^2 + g_1^2}, \label{eq:g_B}\\
g_{BD}&=\frac{\omega_S - \omega_{11}}{2}\sin{2\theta} + g_2\cos{2\theta}. \label{eq:g_BD}
\end{align}

The leakage suppression strategy presented in the main text relies on dynamically tuning the system parameters to enforce the condition $g_{BD}=0$. This nullifies the coupling between the subspace $W_A$ and the dark state $\ket{D}$, effectively confining the gate dynamics to a two-dimensional, leakage-free subspace.

\subsection{Matrix Form of the Controlled-Phase Evolution}
In the main text, we have clarified that our approach is guided by the invariant-subspace principle, with Hamiltonian engineering providing the practical means to block-diagonalize the effective dynamics and thereby suppress spectator-induced leakage by construction. Here, we further connect this invariant-subspace construction to the matrix form of the resulting controlled-phase evolution.

For the single-spectator case discussed in the main text, after the bright-dark-state transformation, the relevant basis can be written as
\begin{equation}
\mathcal{B}
=
\left\{
|00\rangle,
|01\rangle,
|10\rangle,
|11\rangle,
|B\rangle,
|D\rangle，
\right\}.
\end{equation}
When the dynamic block-diagonalization condition \(g_{\mathrm{BD}}=0\) is satisfied, the dark state \(|D\rangle\) is decoupled from the gate dynamics. The only nontrivial evolution involving the logical state \(|11\rangle\) is then confined to the two-dimensional invariant subspace
\begin{equation}
W_{A}
=
\mathrm{span}\left\{
|11\rangle,
|B\rangle
\right\}.
\end{equation}
The effective Hamiltonian in this subspace is
\begin{equation}
H_A
=
\begin{pmatrix}
\omega_{11} & g_B \\
g_B & \omega_B
\end{pmatrix}.
\label{eq:HA_controlled_phase}
\end{equation}
Defining
\begin{equation}
\bar{\omega}
=
\frac{\omega_{11}+\omega_B}{2},
\qquad
\delta
=
\omega_{11}-\omega_B,
\qquad
\Omega
=
\sqrt{g_B^2+\left(\frac{\delta}{2}\right)^2},
\end{equation}
we can write the evolution operator in \(W_A\) as
\begin{equation}
U_A(t)
=
e^{-i\bar{\omega}t}
\begin{pmatrix}
\cos(\Omega t)
-
i\dfrac{\delta}{2\Omega}\sin(\Omega t)
&
-i\dfrac{g_B}{\Omega}\sin(\Omega t)
\\[4pt]
-i\dfrac{g_B}{\Omega}\sin(\Omega t)
&
\cos(\Omega t)
+
i\dfrac{\delta}{2\Omega}\sin(\Omega t)
\end{pmatrix}.
\label{eq:UA_controlled_phase}
\end{equation}

Therefore, in the six-dimensional basis \(\mathcal{B}\), the full physical evolution has the block form
\begin{equation}
U_{\mathrm{phys}}(t)
=
\begin{pmatrix}
e^{i\varphi_{00}(t)} & 0 & 0 & 0 & 0 & 0 \\
0 & e^{i\varphi_{01}(t)} & 0 & 0 & 0 & 0 \\
0 & 0 & e^{i\varphi_{10}(t)} & 0 & 0 & 0 \\
0 & 0 & 0 & u_{11,11}(t) & u_{11,B}(t) & 0 \\
0 & 0 & 0 & u_{B,11}(t) & u_{B,B}(t) & 0 \\
0 & 0 & 0 & 0 & 0 & e^{i\varphi_D(t)}
\end{pmatrix},
\label{eq:Uphys_block_form}
\end{equation}
where the central \(2\times2\) block is precisely \(U_A(t)\),
\begin{equation}
U_A(t)
=
\begin{pmatrix}
u_{11,11}(t) & u_{11,B}(t) \\
u_{B,11}(t) & u_{B,B}(t)
\end{pmatrix}.
\end{equation}
This matrix form explicitly shows that the dark state is excluded from the dynamics and that the \(|11\rangle\)-state evolution is closed within \(W_A\).

A controlled-phase gate is obtained when the evolution in \(W_A\) is cyclic, namely when the population initially in \(|11\rangle\) returns to \(|11\rangle\) at the end of the gate. A simple cyclic condition is
\begin{equation}
\Omega \tau = m\pi,
\qquad
m\in\mathbb{Z}^{+}.
\end{equation}
Under this condition, the off-diagonal elements of \(U_A(\tau)\) vanish:
\begin{equation}
u_{11,B}(\tau)
=
u_{B,11}(\tau)
=
0.
\end{equation}
Thus the state \(|11\rangle\) acquires only a phase,
\begin{equation}
|11\rangle
\rightarrow
e^{i\varphi_{11}}
|11\rangle,
\qquad
e^{i\varphi_{11}}
=
(-1)^m e^{-i\bar{\omega}\tau}.
\end{equation}
The other computational basis states acquire only single-state phases,
\begin{equation}
|ij\rangle
\rightarrow
e^{i\varphi_{ij}}
|ij\rangle,
\qquad
ij\in\{00,01,10\}.
\end{equation}
Projecting the full evolution onto the computational subspace gives
\begin{equation}
U_{\mathrm{comp}}(\tau)
=
\begin{pmatrix}
e^{i\varphi_{00}} & 0 & 0 & 0 \\
0 & e^{i\varphi_{01}} & 0 & 0 \\
0 & 0 & e^{i\varphi_{10}} & 0 \\
0 & 0 & 0 & e^{i\varphi_{11}}
\end{pmatrix}.
\label{eq:Ucomp_phase}
\end{equation}
This evolution can be decomposed as 
\begin{equation} 
U_{\mathrm{comp}}(\tau) =  e^{i\varphi_{00}}
    \left[
    Z_1(\phi_{1})
    \otimes
    Z_2(\phi_{2})
    \right]
    U_{\mathrm{CP}}(\phi_{\mathrm{CP}}),
\label{eq:Ucomp_decomposition} 
\end{equation} 
where $Z(\phi)=\mathrm{diag}(1,e^{i\phi})$, and $U_{\mathrm{CP}}(\phi_{\mathrm{CP}}) = \mathrm{diag}\left( 1, 1, 1, e^{i\phi_{\mathrm{CP}}} \right)$. The corresponding phases are \begin{equation} \phi_1 = \varphi_{10}-\varphi_{00}, \qquad \phi_2 = \varphi_{01}-\varphi_{00}, \qquad \phi_{\mathrm{CP}} = \varphi_{11} - \varphi_{10} - \varphi_{01} + \varphi_{00}. \label{eq:conditional_phase} \end{equation} 
The first factor in Eq.~(\ref{eq:Ucomp_decomposition}) is a global phase, while the next two factors are local single-qubit $Z$ phases. Importantly, these phases are properties of the calibrated gate evolution and are independent of the input quantum state. In experiment, they can be calibrated together with the two-qubit gate and absorbed into the corresponding qubit phase frames through virtual-$Z$ operations, without requiring any knowledge or measurement of the input state. 

After these local phase corrections, the remaining logical operation is $U_{\mathrm{CP}}(\phi_{\mathrm{CP}})$.
The standard controlled-Z gate is obtained as the special case
\begin{equation}
\phi_{\mathrm{CP}}=\pi,
\qquad
U_{\mathrm{CZ}}
=
\mathrm{diag}(1,1,1,-1).
\label{eq:UCZ_standard}
\end{equation}

This derivation shows that the role of invariant-subspace engineering is to make the physical multi-level evolution reduce to a closed logical controlled-phase operation. The standard CZ gate demonstrated in the main text is therefore a special case of a more general controlled-phase construction based on block-diagonal Hamiltonian engineering.

The phase acquired by \(|11\rangle\) during the closed excursion in \(W_A\) also has a geometric interpretation. In the effective two-level system \(\{|11\rangle,|B\rangle\}\), the traceless part of \(H_A\) generates a closed trajectory on the Bloch sphere when the cyclic condition is satisfied. The associated nonadiabatic geometric phase is determined by the solid angle enclosed by this trajectory. In the resonant case, \(\delta=0\), the dynamical phase from the traceless Hamiltonian vanishes, and the phase acquired after a closed \(2\pi m\) rotation is geometric. In the detuned case, the observed controlled phase generally contains both geometric and dynamical contributions, while the single-qubit and global phases can be calibrated or absorbed into \(Z\)-frame updates. This closed-subspace picture naturally supports both the standard CZ gate and more general controlled-phase gates.

\subsection{Block-Diagonalization for General Multi-Level Systems}

The block-diagonalization strategy presented for a three-level system can be generalized to more complex scenarios involving multiple spectator levels $\{S_1, S_2, ..., S_n\}$. Here, we provide a rigorous derivation based on the concept of an invariant subspace.

Consider a system where the gate states $\{\ket{11}, \ket{02}\}$ interact with $n$ spectator levels. The dynamics are confined to an $(n+2)$-dimensional Hilbert space $W = \{\ket{11}, \ket{02}, \ket{S_1}, ..., \ket{S_n}\}$, with the general effective Hamiltonian:
\begin{align}
    H_n^{\text{eff}} = \sum_{i} \omega_i \ket{i}\bra{i} + \sum_{i \neq j} \left( g_{ij} \ket{i}\bra{j} + \text{h.c.} \right),
    \label{eq:H_n_eff}
\end{align}
where the indices $i, j$ run over the entire basis set. Our goal is to find a two-dimensional subspace $W_A \subset W$ that is invariant under the action of $H_n^{\text{eff}}$ and contains the initial state $\ket{11}$. If such a subspace exists, its orthogonal complement is also invariant, leading to a natural block-diagonalization of the Hamiltonian.

The derivation proceeds in two steps. First, we determine the basis of the invariant subspace, $W_A = \{\ket{11}, \ket{B}\}$, where $\ket{B}$ is orthogonal to $\ket{11}$. The invariance condition $H_n^{\text{eff}}\ket{11} \in W_A$ implies that the component of $H_n^{\text{eff}}\ket{11}$ orthogonal to $\ket{11}$ must be proportional to $\ket{B}$. This uniquely defines the form of the ``generalized bright state'' as:
\begin{align}
    \ket{B} = \frac{1}{g_B} \sum_{j \neq 11} g_{11,j} \ket{j},
    \label{eq:B_generalized}
\end{align}
where $g_B = \sqrt{\sum_{j \neq 11} |g_{11,j}|^2}$ is a normalization factor that becomes the new effective gate coupling.

Second, we impose the closure condition $H_n^{\text{eff}}\ket{B} \in W_A$. This condition implies that $H_n^{\text{eff}}\ket{B}$ must be a linear combination of $\ket{11}$ and $\ket{B}$. We can write this as:
\begin{align}
    H_n^{\text{eff}}\ket{B} = c_1\ket{11} + c_2\ket{B},
\end{align}
where $c_1$ and $c_2$ are some coefficients. From the first step of our derivation, we know $\braket{B|H_n^{\text{eff}}|11} = g_B$. Therefore, $c_1 = g_B$. This simplifies the condition to finding a unique scalar $\omega_B$ such that:
\begin{align}
\label{eq:bright_state_eigen_equation}
    H_n^{\text{eff}}\ket{B} = g_B\ket{11} + \omega_B\ket{B}.
\end{align}
This equation is not automatically satisfied and imposes a set of strong constraints on the coupling parameters. Specifically, it demands that the state $(H_n^{\text{eff}}\ket{B} - g_B\ket{11})$ must be perfectly proportional to $\ket{B}$. This leads to $n + 1$ independent constraint equations. If the system possesses at least $n$ tunable parameters (e.g., $n$ independently tunable couplers), these constraints can, in principle, be satisfied, leading to a successful block-diagonalization.

When all conditions are met, the Hamiltonian is successfully block-diagonalized, and the dynamics of $\ket{11}$ are strictly confined to the invariant subspace $W_A$. The evolution is then governed by the simple $2 \times 2$ Hamiltonian:
\begin{align}
    H_A = 
    \begin{pmatrix}
    \omega_{11} & g_B  \\
    g_B & \omega_{B}
    \end{pmatrix},
\end{align}
where $\omega_B = \braket{B|H_n^{\text{eff}}|B}$ is the energy of the generalized bright state. In conclusion, we have generalized our block-diagonalization scheme to arbitrary multi-level systems. While not always achievable due to physical constraints on tunability, this invariant subspace method provides a powerful and systematic framework for designing leakage-elimination protocols in complex quantum systems.

Let us now apply this general framework to the specific and experimentally relevant case of multiple spectator qubits. As a practical and often valid approximation, we assume that the direct interactions between different spectator qubits are negligible higher-order effects. This simplification, while not essential for the block-diagonalization principle itself, allows for a more tractable analysis. Under this assumption, the effective Hamiltonian for a system with $n$ spectator qubits interacting with the gate states $\{\ket{11}, \ket{02}\}$ is:
\begin{align}
    H_n^{\text{eff}} = &  \omega_{11}\ket{11}\bra{11} + \omega_{02}\ket{02}\bra{02} + g_{\mathrm{gate}}(\ket{11}\bra{02} + \text{h.c.})  \notag \\
    & + \sum_{i=1}^n \left( \omega_{S_i}\ket{S_i}\bra{S_i} + g_{1,i}(\ket{11}\bra{S_i} + \text{h.c.}) + g_{2,i}(\ket{02}\bra{S_i} + \text{h.c.}) \right),
\end{align}
where $g_{1,i} = \braket{11|H_n^{\text{eff}}|S_i}$ and $g_{2,i} = \braket{02|H_n^{\text{eff}}|S_i}$.

For this multi-spectator Hamiltonian, the closure condition from Eq.~\eqref{eq:bright_state_eigen_equation} imposes a set of constraints that can be expressed as:
\begin{align}
\label{eq:constraint_multi_1}
\sum_{i=1}^n g_{1,i} g_{2,i} + g_{\mathrm{gate}} \omega_{02}&= g_{\mathrm{gate}}\omega_B, \\
\label{eq:constraint_multi_2}
g_{\mathrm{gate}} g_{2,i} + g_{1,i} \omega_{11} &= g_{1,i} \omega_B ,
\end{align}
for each spectator $i \in \{1, ..., n\}$. Here, $\omega_B$ is the unique energy of the generalized bright state. Let us consider the common experimental scenario where the couplings $g_{1,j}$ and fixed, while the couplings $g_{2,j}$ are tunable . To satisfy the above system of equations, the tunable couplings $g_{2,j}$ must be set to:
\begin{align}
\label{eq:g2_solution}
g_{2,i} = \frac{g_{1,i} \left( \sum_{j \neq i} g_{1,j}^2 (\omega_{S_j} - \omega_{S_i}) +  g_{\mathrm{gate}}^2 (\omega_{S_i} - \omega_{11}) \right)}{g_{\mathrm{gate}} \left( g_{B}^2 - g_{1,i}^2 \right)},
\end{align}
where $g_B^2 = g_{\mathrm{gate}}^2 + \sum_k g_{1,k}^2$.

A crucial simplification occurs in the weak coupling limit, where $g_{1,j} \ll g_{\mathrm{gate}}$ for all $j$. In this regime, the first term in the numerator of Eq.~\eqref{eq:g2_solution} becomes negligible compared to the second, and the denominator approximates to $g_{\mathrm{gate}}^3$. The condition for $g_{2,j}$ then reduces to:
\begin{align}
g_{2,i} \approx \frac{g_{1,j}(\omega_{S_j} - \omega_{11})}{g_{\mathrm{gate}}}.
\end{align}
This expression is identical to the suppression condition for a single spectator in a three-level system. It explains our key experimental observation: in a multi-spectator scenario, the optimal setting for each spectator coupler is independent of the state and parameters of the other spectators, thus confirming the scalability of our protocol.

\subsection{Summary of Leakage Suppression under Different Configurations}

The effectiveness of our block-diagonalization scheme hinges on the ability to dynamically control the leakage couplings, $g_1$ and $g_2$, to satisfy the $g_{BD}=0$ condition. This control is determined by the system's physical topology and the specific resonance condition. We summarize the applicability for several common configurations in Table~\ref{tab:leakage_summary}.
\begin{table}[htbp]
  \centering
  \caption{Summary of leakage suppression capability for different topologies and resonance conditions. The parameters $g_1 = \bra{S}H_{\text{eff}}\ket{11}$ and $g_2 = \bra{S}H_{\text{eff}}\ket{02}$ depend on the specific physical implementation.}
  \label{tab:leakage_summary}
  \renewcommand{\arraystretch}{2}
  \setlength{\tabcolsep}{10pt}

  \begin{tabular}{ccccc}
    \toprule
    \textbf{Topology}
    & \shortstack[c]{\textbf{Resonance}\\\textbf{Condition}}
    & \shortstack[c]{\textbf{Spectator}\\\textbf{State $\ket{S}$}}
    & \shortstack[c]{\textbf{Coupling Strengths} \textbf{($g_1, g_2$)}}
    & \shortstack[c]{\textbf{Suppression}\\\textbf{Possible?}} \\
    \midrule

    l-h-s
    & $\omega_s \approx \omega_l$
    & $\ket{011}$
    & \parbox[c]{4.8cm}{\centering
        $g_1 = g_{ls}^{01},\;
        g_2 = g_{hs}^{20}\!\left(\omega_{c_s}\right)$}
    & Yes \\

    l-h-s
    & $\omega_s \approx \omega_h$
    & $\ket{101}$
    & \parbox[c]{4.8cm}{\centering
        $g_1 = g_{hs}^{01}\!\left(\omega_{c_s}\right),\;
        g_2 \approx 0$}
    & Yes \\

    h-l-s
    & $\omega_s \approx \omega_l$
    & $\ket{011}$
    & \parbox[c]{4.8cm}{\centering
        $g_1 = g_{ls}^{01}\!\left(\omega_{c_s}\right),\;
        g_2 = g_{hs}^{20}$}
    & Yes \\

    h-l-s
    & $\omega_s \approx \omega_h$
    & $\ket{101}$
    & \parbox[c]{4.8cm}{\centering
        $g_1 = g_{hs}^{01},\;
        g_2 \approx 0$}
    & Challenging / No \\
    \bottomrule
  \end{tabular}
\end{table}
As shown in the first three scenarios of the table, our method is effective when at least one of the key leakage couplings ($g_1$ or $g_2$) is strongly dependent on the tunable spectator coupler frequency, $\omega_{c_s}$. This tunability provides the necessary degree of freedom to enforce the $g_{BD}=0$ condition and suppress leakage.

However, a challenging case arises, for instance, in the h-l-s topology with $\omega_s \approx \omega_h$. Here, the dominant leakage pathway coupling $\ket{110}$ and $\ket{S}=\ket{101}$ is often a direct, fixed stray capacitance between $Q_h$ and $Q_s$. If this fixed coupling dominates $g_1$, and the tunable coupler does not significantly mediate this interaction, our method lacks sufficient control to nullify $g_{BD}$, leading to persistent leakage.

This highlights a key aspect of our approach: it is designed to eliminate leakage pathways that are mediated by tunable elements. Leakage dominated by fixed stray couplings falls outside the primary scope of this technique and typically requires mitigation through other means, such as optimized chip design or careful frequency allocation. Therefore, our dynamic control scheme is a powerful, complementary tool that addresses a significant class of spectator errors, thereby expanding the parameter space for high-fidelity operations in scalable quantum processors.

\subsection{State-Dependent Leakage Channels and Generalized Suppression Strategy}
\label{sec:supp_statedependent}
\begin{figure}[h]
    \centering
    \includegraphics[width=14.2cm]{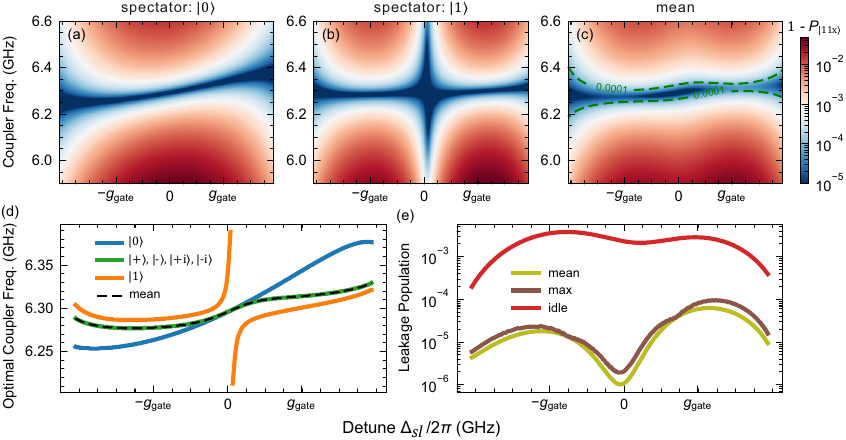}
    \caption{
    State-dependent leakage and generalized suppression.
    (a) Simulated leakage population as a function of coupler frequency and spectator-gate detuning $\Delta_{sl}$ for a spectator initialized in the $\ket{0}$ state.
    (b) Same as (a), but for a spectator initialized in the $\ket{1}$ state.
    (c) Leakage averaged over six cardinal spectator states ($\ket{0}, \ket{1}, \ket{\pm}, \ket{\pm i}$). The dashed green lines indicate the $10^{-4}$ infidelity contour.
    (d) Optimal coupler frequency as a function of detuning for different initial spectator states. The blue and orange lines correspond to the $\ket{0}$ and $\ket{1}$ states, respectively. The green line is for the four superposition states, and the black dashed line is the optimal frequency extracted from the mean infidelity in (c).
    (e) Leakage population versus detuning. The ``idle'' curve shows the mean leakage with the coupler at a fixed, non-optimal frequency. The ``mean'' and ``max'' curves show the mean and maximum leakage among the six cardinal states when operating at the optimal coupler frequency defined by the black dashed line in (d).}
    \label{fig:supp_leakage}
\end{figure}
The theoretical framework presented in the main text is general. However, the physical realization of the three-level system, particularly the leakage state $\ket{S}$ and the couplings $g_1$ and $g_2$, depends on the initial state of the spectator qubit. In the main text, we focused on the case where the spectator is in its ground state $\ket{0}$. Here, we analyze the case of an excited spectator and propose a generalized suppression strategy for a spectator in an arbitrary superposition state.

First, we consider a spectator initialized in the excited state, $\ket{1}$. The relevant gate-level states are now $\ket{111}$ and $\ket{021}$, and the primary leakage state becomes $\ket{S} = \ket{120}$. This forms a new three-level system that can still be described by the general Hamiltonian in Eq.~(1) of the main text. However, the roles of the direct and coupler-mediated interactions are swapped. The effective leakage couplings become $g_1 = g_{hs}^{20}$ and $g_2 = g_{ls}^{01}$. The block-diagonalization condition, $g_{BD} = 0$, remains the necessary condition to eliminate leakage. Fig.~\ref{fig:supp_leakage}(b) shows the simulated leakage for this scenario, confirming that a ``valley'' of near-zero leakage exists where the coupler frequency can be tuned to completely suppress this leakage channel.

However, a comparison of the leakage maps for the ground-state [Fig.~\ref{fig:supp_leakage}(a)] and excited-state [Fig.~\ref{fig:supp_leakage}(b)] spectators reveals that their optimal suppression frequencies are markedly different. This poses a challenge when the spectator is in an arbitrary superposition state, $\ket{\psi}_s = \alpha\ket{0} + \beta\ket{1}$, as both leakage channels are active simultaneously. To investigate this, we simulate the gate performance for spectators initialized in six cardinal states: $\ket{0}$, $\ket{1}$, and the superposition states $\left(\ket{0} + \ket{1}\right)/\sqrt{2}$, $\left(\ket{0} - \ket{1}\right)/\sqrt{2}$, $\left(\ket{0} + i\ket{1}\right)/\sqrt{2}$, and $\left(\ket{0} - i\ket{1}\right)/\sqrt{2}$. We assess the performance by the final population remaining in the $\ket{11x}$ subspace, $P_{\ket{11x}} = P_{\ket{110}} + P_{\ket{111}}$.

Fig.~\ref{fig:supp_leakage}(c) shows the leakage population ($1 - P_{11x}$) averaged over these six initial states. While the leakage can no longer be completely eliminated for all states simultaneously, a clear optimal coupler frequency exists that minimizes this average infidelity. We extract the optimal coupler frequency that minimizes leakage for each scenario, as shown in Fig.~\ref{fig:supp_leakage}(d). Notably, the optimal frequencies for all superposition states are identical and align perfectly with the optimal frequency for the mean-infidelity case. This occurs because the total leakage is, to a good approximation, a weighted average of the two independent channels, $P_{\text{leak}} = |\alpha|^2 P_{\text{leak},0} + |\beta|^2 P_{\text{leak},1}$.

This result suggests a practical and robust solution: use the mean optimal frequency as a universal suppression setting. To evaluate its effectiveness, we analyze the leakage while operating at this frequency, as shown in Fig.~\ref{fig:supp_leakage}(e). Compared to the idle-coupler case, the mean leakage is suppressed by more than an order of magnitude, remaining below $10^{-4}$. Furthermore, we consider the worst-case scenario. The ``max'' curve, representing the maximum leakage among the six initial states, is also robustly suppressed below $10^{-4}$. These findings validate that our generalized suppression strategy is both effective and robust for protecting a spectator qubit in an arbitrary quantum state.

\section{Device and Experimental Setup}
\begin{figure}[htbp]

    \centering
    \includegraphics[width=0.6\linewidth]{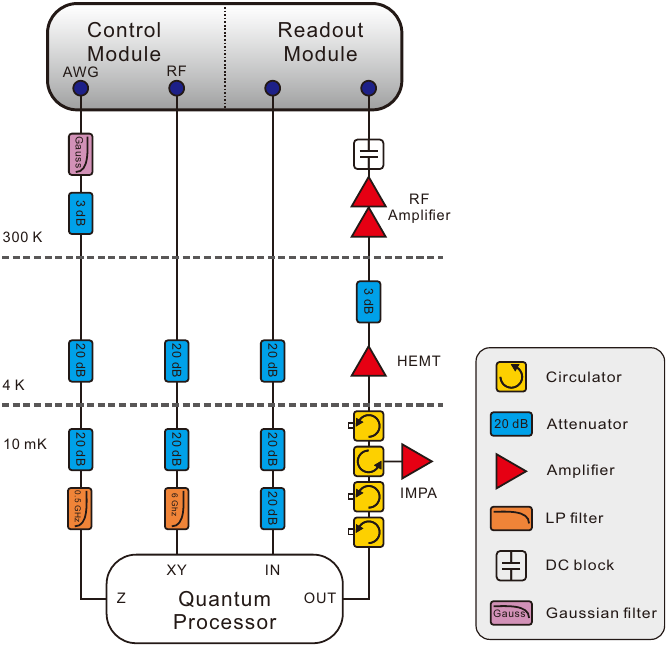}
    \caption{Details of wiring and experimental setup}
    \label{fig:wiring}

\end{figure}
\subsection{Chip Device}

The experiments were performed on the \textit{Wukong} processor, a large-scale quantum chip integrating 72 qubits and 126 tunable couplers. The device employs a multi-layer, flip-chip architecture, which physically separates the qubit layer from the routing and control layer. This design choice is crucial for achieving high-density interconnects while effectively mitigating signal crosstalk and parasitic coupling, further suppressed by optimized air bridges and other shielding structures.

The fundamental building block of the processor is a qubit-coupler-qubit (QCQ) unit, designed for high-fidelity two-qubit operations. A unique feature of our design is a waveguide resonator placed between adjacent qubits, which facilitates a direct, capacitive qubit-qubit coupling ($g_{qq}$). This allows for a larger physical separation between qubits, which not only mitigates unwanted XY crosstalk but also provides essential space for routing control and readout lines without compromising connectivity.

Both the qubits and the couplers are implemented as flux-tunable transmon qubits with asymmetric SQUID loops. This specific design offers a practical balance between achieving wide frequency tunability, which is essential for avoiding frequency collisions, and maintaining resilience to magnetic flux noise. Each qubit is equipped with a full set of independent control lines (XY for rotation, Z for frequency tuning) and a dedicated readout resonator, ensuring precise state manipulation and high-fidelity measurement.

Crucially for this work, each tunable coupler is also controlled by a dedicated Z-flux line. This individual addressability allows for the application of fast, dynamic flux pulses to the couplers, which is the core mechanism enabling our leakage suppression protocol. The drive and readout of the couplers themselves, when needed for characterization, are facilitated through a probe-drive mechanism. This combination of high connectivity, low crosstalk, and full individual control over every element provides a robust and highly scalable platform for implementing advanced, dynamic multi-qubit gate protocols like the one presented in this study.

\subsection{Measurement Setup}

The experiment is conducted in a Bluefors XL1000 dilution refrigerator, providing an ultra-low temperature environment necessary for quantum coherence. Magnetic shielding is achieved using a permalloy shield placed at the mixing chamber stage, which effectively reduces external magnetic interference that could otherwise degrade qubit coherence and measurement accuracy.

The qubit control and readout signals are generated through a dual-stage down-conversion process, where local oscillators and arbitrary waveform generators (AWGs) provide the necessary signals as depicted in Fig.~\ref{fig:wiring}. For qubit manipulation, the XY control pulses are generated by the AWG and upconverted to the qubit frequency using mixers. The flux control for the qubits is also generated by the AWG, which produces both static bias voltages and dynamic waveforms. These signals are synthesized and output as AC signals directly from the Tianji control system, ensuring precise flux control over the qubits.

The readout resonators are coupled to a transmission line, which routes the signals to an impedance-matched parametric amplifier (IMPA) for amplification~\cite{2021_impa}. The IMPA is pumped and biased using a microwave source and a DC source, respectively, providing more than 15 dB of gain across a 500 MHz bandwidth. This ensures a high signal-to-noise ratio, allowing for accurate qubit state readout.

All signals, including control and readout, are generated by the Tianji integrated system, which provides synchronized waveform generation for both qubit control and measurement. The signals are routed through a series of attenuators at various temperature stages within the cryostat before reaching the chip, minimizing thermal noise and ensuring the integrity of the signals.

\subsection{Chip Parameters and Characterization}

\begin{table*}[tbh]
\caption{Device parameters. 
$\omega_r$ is the readout resonator frequency. $\omega_{\mathrm{idle}}$ is $\ket{0} \rightarrow \ket{1}$ transition frequency at the idle configuration where the residual ZZ coupling is minimized. 
$\alpha/2\pi$ is the anharmonicity. 
$T_1$, $T_2^*$ and $T_2^{\mathrm{echo}}$ are the energy decay time, ramsey dephasing time, and spin-echo dephasing time measured at the idle configuration. $F_0$($F_1$) is the readout fidelity of $\ket{0}$($\ket{1}$). 
$e_1$ is the single-qubit gate error measured via simultaneous Randomized Benchmarking (RB). 
$\rho_{1c}$ and $\rho_{2c}$ are the frequency-independent coupling coefficients between the qubit and coupler.  
$\rho_{qc}$ is the frequency-independent direct coupling coefficient between the qubit and qubit.}
\begin{threeparttable}
\begin{tabular*}{\textwidth}{@{\extracolsep{\fill}}cccccccc}
\toprule
 &{$Q_l$} &{$Q_h$} &{$Q_s(Q_{s_{1}})$} &{$Q_{s_{2}}$} &{$Q_{s_{3}}$} &{$C_g$} &{$C_s$} \\
\midrule
$\omega_r/2\pi$ (GHz) &{$7.214$} &{$7.051$} &{$7.132$} &{$7.268$} &{$7.176$} &{$\sim$} &{$\sim$} \\
$\omega_{\mathrm{idle}}/2\pi$ (GHz) &{$4.26$} &{$4.624$} &{$4.147$} &{$4.075$} &{$4.215$} &{$6.384$} &{$6.406$} \\
$\alpha/2\pi$ (MHz) &{$-233$} &{$-236$} &{$-247$} &{$-244$} &{$-241$} &{$-119$} &{$-120$} \\
$T_1$ ($\mu$s) &{$24.1$} &{$14.1$} &{$14.8$} &{$13.7$} &{$23.4$} &{$\sim$} &{$\sim$} \\
$T_2^*$ ($\mu$s) &{$2.0$} &{$2.1$} &{$2.5$} &{$1.9$} &{$1.9$} &{$\sim$} &{$\sim$} \\
$T_2^{\mathrm{echo}}$ ($\mu$s) &{$7.9$} &{$5.7$} &{$7.0$} &{$5.2$} &{$8.8$} &{$\sim$} &{$\sim$} \\
$F_0$ (\%) &{$94.1$} &{$96.1$} &{$92.0$} &{$93.9$} &{$95.5$} &{$\sim$} &{$\sim$} \\
$F_1$ (\%) &{$88.4$} &{$88.5$} &{$87.7$} &{$88.8$} &{$83.7$} &{$\sim$} &{$\sim$} \\
$e_1$ &{$0.22$} &{$0.21$} &{$0.26$} &{$0.31$} &{$0.24$} &{$\sim$} &{$\sim$} \\
$\rho_{1c}$ &{$\sim$} &{$\sim$} &{$\sim$} &{$\sim$} &{$\sim$} &{$0.0223$} &{$0.0241$} \\
$\rho_{2c}$ &{$\sim$} &{$\sim$} &{$\sim$} &{$\sim$} &{$\sim$} &{$0.0249$} &{$0.0249$} \\
$\rho_{qq}$ &{$\sim$} &{$\sim$} &{$\sim$} &{$\sim$} &{$\sim$} &{$0.0029$} &{$0.0029$} \\
\bottomrule
\end{tabular*}
\label{Table: device params}
\end{threeparttable}
\end{table*}

To mitigate frequency collisions in large-scale quantum chips, the operating frequencies for all 65 single-qubit gates and 90 two-qubit gates were assigned using a snake-like frequency allocation scheme~\cite{2024_freq_allo_google}. This allocation process takes into account various potential collisions ensuring optimal distribution of qubit and coupler frequencies. For the experiments, we selected 5 qubits and 4 couplers from this allocation, with their respective single-qubit and two-qubit operating points following the frequency allocation results. The specific parameters of these qubits are listed in Table.~\ref{Table: device params}.

The physical parameters of the gate coupler and one of the spectator couplers were extracted using the method~\cite{2023_Character_Coupler}. The primary approach involved using one neighboring qubit as a drive and the other as a dispersive readout to characterize the coupler properties. The qubit-qubit coupling strength ($g_{qq}$) and qubit-coupler coupling strength ($g_{qc}$) were determined through swap experiments, providing accurate measurements of the interaction strengths within the system.

For single-qubit gates, we used a 30 ns DRAG pulse envelope, with calibration following standard methods from previous works. The two-qubit CZ gate was implemented with a 40 ns flat-top Gaussian waveform, featuring 1.25 ns sigma for the qubits, 2 ns sigma for the coupler, and a 7.5 ns buffer for both. Calibration procedures followed established references to ensure optimal performance and minimal leakage~\cite{2021_CZ_iswap_zzfree}.

\subsection{Flux Crosstalk Characterization}

\begin{figure}[tbh]

    \centering
    \includegraphics[width=0.83\linewidth]{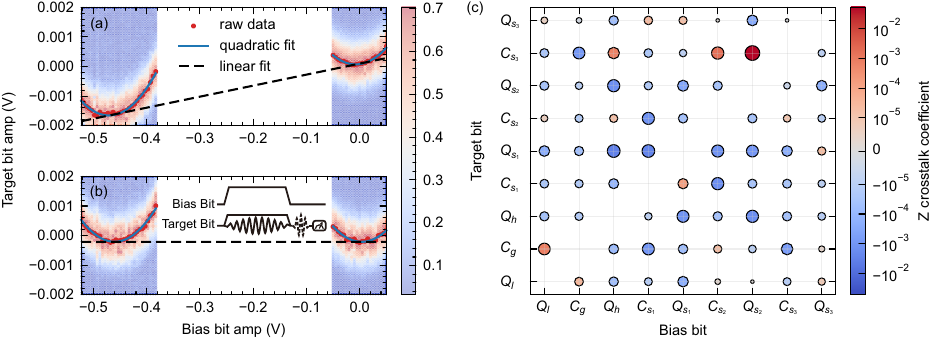}
    \caption{Z Crosstalk Characterization and Compensation in a 5-Qubit, 4-Coupler System. (a) (b) Heatmap of bias Z amplitude versus target Z amplitude before and after Z crosstalk correction, with raw peak P1 probabilities (red points), quadratic fitting (blue line), and linear fitting (black dashed line) to extract the crosstalk coefficient. Inset: Pulse sequence used for crosstalk measurement, featuring a Z pulse on the bias qubit and a compensatory Z pulse on the target qubit. (c) Crosstalk matrix heatmap, where circle size and color indicate the magnitude of coefficients $M_{i,j}$ (range: $10^{-5}$ to $10^{-2}$).
}
    \label{figS3}

\end{figure}

In order to accurately control the frequencies of qubits and couplers in a multi-qubit system, particularly when exploring interactions involving multiple spectator qubits, it is essential to address flux crosstalk between Z control lines. The presence of crosstalk complicates the precise tuning of qubits and couplers, requiring careful calibration of all Z lines.

Our method for measuring crosstalk involves applying a Z pulse with a specific amplitude (Zamp) to the bias qubit and compensating for the resulting frequency shift on the target qubit by applying a small Zamp to its Z line. When the compensation is correct, the frequency of the target qubit remains unchanged, allowing for successful excitation by a $\pi$ pulse. By scanning the Zamp values for both qubits, we can map their relationship. For non-neighboring qubits or couplers, the relationship between the two Zamp values is linear, and the crosstalk coefficient can be extracted through linear fitting. However, for neighboring qubits and couplers, where the coupling exceeds 100 MHz, the Zamp relationship becomes strongly nonlinear. Typically, this nonlinear behavior is modeled as a combination of coupling terms and linear crosstalk terms, but this approach often lacks precision~\cite{2023_Character_Coupler}.

To improve accuracy, we present a highly precise method using the coupler-qubit crosstalk as an example. First, we measure the voltage corresponding to the coupler's peak frequency and its modulation period. Near the two closest peak points, we scan the Zamp relationship and perform a quadratic fit to extract the peak positions. A linear fit of these peak positions then yields the linear crosstalk coefficient between the coupler and the neighboring qubit in Fig.~\ref{figS3}(a). This method effectively isolates the classical crosstalk by exploiting the fact that the coupler exerts the same frequency ``push'' on the qubit at identical frequencies, thereby eliminating the nonlinear components from the voltage relationship. This approach provides two orders of magnitude higher accuracy compared to traditional nonlinear fitting methods.

The crosstalk matrix for the 5 qubits and 4 couplers used in the experiment is shown in Fig.~\ref{figS3}(b). Most of the crosstalk coefficients are below 1e-3, with only a few notable exceptions. During compensation, the required voltage amplitudes can be calculated as 
\begin{equation}
	\mathbf{z}_{\mathrm{corrected}}=\mathbf{M}^{-1}\mathbf{z},
\end{equation}
and sent to the control system.

\section{Additional Characterization and Analysis of Spectator-Induced Leakage}

\subsection{Characterization of the Effective Three-Level Hamiltonian}
\begin{figure}
    \centering
    \includegraphics[width=0.5\linewidth]{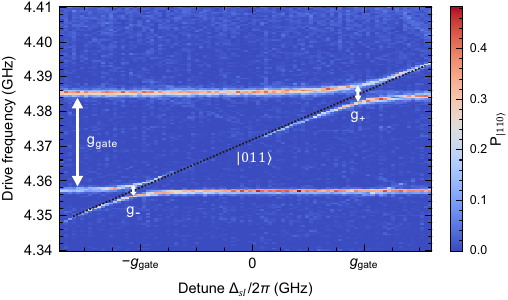}
    \caption{Spectroscopy of the three-level system. The population of the gate-subspace state $\ket{110}$ is plotted as a function of the drive frequency and the detuning of the spectator qubit $\Delta_{sl}$. The two horizontal lines are the hybridized gate levels, separated by $g_{\mathrm{gate}}$. The diagonal line tracks the spectator state $\ket{011}$. Two avoided crossings with energy gaps $g_+$ and $g_-$ are visible, indicating the coupling between the spectator and the gate subspace. A two-photon drive was used, so the drive frequency corresponds to half the actual energy level transition.
}
    \label{fig:engegy_spectrum}
\end{figure}
To provide experimentally grounded parameters for our theoretical model, we characterized the key coupling strengths of the effective three-level Hamiltonian using spectroscopy. The experiment was performed on the l-h-s topology with the spectator state $\ket{S} = \ket{011}$. To enhance the visibility of the coupling, the spectator coupler was tuned to $\omega_{c_s}/2\pi = 5.5$~GHz, thereby increasing the interaction strength.

The experimental procedure involves setting the gate states $\ket{110}$ and $\ket{020}$ to be resonant. We then apply a two-photon drive tone while sweeping its frequency and the spectator qubit frequency. The resulting spectrum is shown in Fig.~\ref{fig:engegy_spectrum}. The two visible avoided crossings are direct signatures of the interaction between the spectator state and the hybridized gate levels. The energy gaps at these anti-crossings, labeled $g_+$ and $g_-$, correspond to the coupling strengths of $\ket{011}$ to the symmetric, $\frac{1}{\sqrt{2}}(\ket{020} + \ket{110})$, and antisymmetric, $\frac{1}{\sqrt{2}}(\ket{020} - \ket{110})$, superpositions of the gate states, respectively.

These experimentally measured gaps are directly related to the coupling parameters $g_1 = \langle 110 | H_{\mathrm{eff}} | 011 \rangle$ and $g_2 = \langle 020 | H_{\mathrm{eff}} | 011 \rangle$ in the original basis via the transformation:
\begin{align}
    g_1 &= \frac{g_+ + g_-}{2}, \\
    g_2 &= \frac{g_+ - g_-}{2}.
\end{align}
By fitting the energy gaps in the spectrum, we extract the values for $g_+$ and $g_-$, and subsequently calculate $g_1$ and $g_2$. These parameters serve as experimentally-validated inputs for the simulations presented in the main text.

\subsection{Leakage Amplification Sequence Based on Interference}

\begin{figure}[h]
    \centering
    \includegraphics[width=0.8\linewidth]{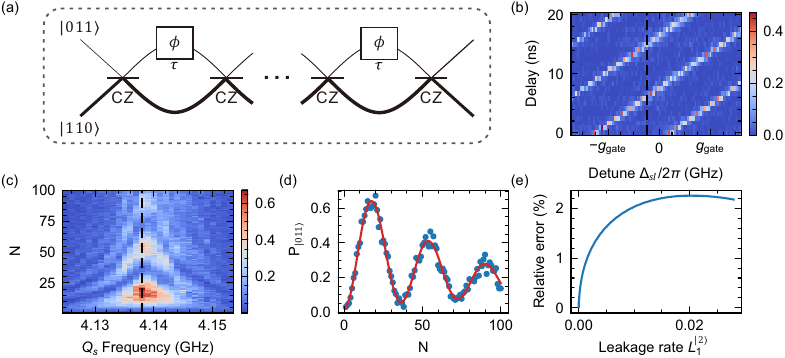}
    \caption{(a) Schematic representation of the CZ leakage amplification sequence, conceptualized as an interferometer where the CZ gate functions as a beam splitter. The phase difference $\phi$ between the interferometer arms is independently adjustable via the parameters $\tau$ (delay time) and $\Delta$ (detuning). (b) Coarse sweep of the delay time, unveiling equidistant interference fringes in the leakage amplification signal across a range of spectator qubit frequencies. (c) Fine adjustment of the interferometer arm phase by varying the $Q_l$ oscillation within the delay time. (d) Leakage signal along the slowest oscillating black dashed line in panel (c), with experimental data (blue points) and fitted theoretical model (red curve) as a function of the CZ gate number. (e) Theoretical relative error on the extracted leakage rate as a function of parasitic leakage to $\ket{020}$, justifying the two-level approximation.}
    \label{fig:leak_floquet}
\end{figure}

To precisely measure small leakage rates, we implemented a sequence of $ N $ consecutive controlled-Z (CZ) gates and measured the population of the spectator qubit state. Each CZ gate acts as a Landau-Zener (LZ) transition, serving as a beam splitter, but successive LZ transitions result in a slow accumulation of population. To accelerate this process, we introduced idle periods $\tau$ between consecutive CZ gates, analogous to the arms of a Mach-Zehnder (MZ) interferometer, as schematically shown in Fig.~\ref{fig:leak_floquet}(a). These idle periods allow the accumulation of a relative phase between the $\ket{110}$ and $ |011\rangle $ states, which amplifies the population transfer between these states. The process is analyzed using an advanced Floquet model, but here we focus on the leakage from the initial state $ |110\rangle $ to $ |011\rangle$, which can be described using a simplified SU(2) formalism.

An arbitrary SU(2) rotation can be expressed as:
\begin{equation}
    R(\beta,\zeta,\chi) = \begin{pmatrix} 
    e^{-i\zeta}\cos\frac{\beta}{2} & -ie^{i\chi}\sin\frac{\beta}{2} \\
    -ie^{-i\chi}\sin\frac{\beta}{2} & e^{i\zeta}\cos\frac{\beta}{2} 
    \end{pmatrix}
    = Z(\zeta-\chi)RX(\beta)Z(\zeta+\chi),
\end{equation}
where $\chi$ has no measurable effect, and is set to zero. The rotation matrix simplifies to:
\begin{equation}
    R(\beta, \zeta) = \begin{pmatrix} 
    e^{-i\zeta} \cos\beta & -i \sin\beta \\
    -i \sin\beta & e^{i\zeta} \cos\beta 
    \end{pmatrix}
    = e^{-i\Omega \sigma_n},
\end{equation}
which describes a rotation by angle $ \Omega $ around an axis $ \mathbf{\sigma_n} $ in the XZ plane, where $ \Omega = \arccos\left(\cos\beta \cos\zeta\right) $, and
\begin{equation}
    \mathbf{\sigma_n} = \begin{pmatrix}
    \cos\alpha & \sin\alpha \\
    \sin\alpha & -\cos\alpha
    \end{pmatrix},
    \quad \alpha = \arctan\left( \frac{\tan\beta}{\sin\zeta} \right).
\end{equation}
After $ N $ CZ gates, the probability of measuring the $ |011\rangle $ state is:
\begin{equation}
    P_{\ket{011}} = \sin^2\alpha \sin^2 N\Omega.
\end{equation}

To further amplify the leakage, we introduce idle periods, modeled as $ Z(\phi) $ gates, where $ \phi = \Delta_{sl}^{\mathrm{idle}}\tau $. The parameter $ \zeta $ in the original CZ gate is modified to $ \zeta' = \zeta + \frac{\phi}{2} $. By adjusting the duration $\tau$ of the idle periods, we fit the relationship between $N$ and $ P_{011}$ and extract the leakage probability per CZ gate for the specific initial state $\ket{110}$, which is $P(\ket{110} \to \ket{011}) = \sin^2(\beta)$. To relate this to the average leakage rate $L_1$ used in XEB, we must consider that the CZ gate only acts non-trivially on the $\ket{11}$ state. The other three computational basis states ($\ket{00}, \ket{01}, \ket{10}$) ideally do not couple to the spectator state and thus do not contribute to this leakage channel. Therefore, the leakage probability averaged over the entire four-dimensional computational subspace is one-fourth of the leakage from the $\ket{11}$ state alone. This leads to the relation:
\begin{equation}
    L_1 = \frac{1}{4} P(\ket{110} \to \ket{011}) = \frac{\sin^2\beta}{4}.
\end{equation}

To achieve maximum oscillation amplitude and thus the most accurate extraction of $L_1$, the effective phase must be set to $\zeta' = \pi/2$. This requires precise control over the accumulated phase $\phi$. However, the delay time $\tau$ is limited by the sampling rate of our arbitrary waveform generator. To overcome this, we instead finely tune the spectator qubit's frequency $\omega_s^{\mathrm{idle}}$ during the delay period, which allows for precise, continuous control over the detuning $\Delta_{sl}^{\mathrm{idle}}$ and thus the phase $\phi$.

The experimental procedure is as follows. First, by scanning the spectator qubit frequency and applying the above procedure, we obtain Fig.~\ref{fig:leak_floquet}(b), showing evenly spaced interference fringes—a hallmark of the MZ interferometer. For each spectator qubit frequency, we identified the optimal delay that maximizes constructive interference. Then, with $\tau$ fixed, we finely sweep $\omega_s^{\mathrm{idle}}$ to map out the interference signal versus the number of cycles $N$, as shown in Fig.~\ref{fig:leak_floquet}(c). The central bright fringe indicates the optimal phase condition. A slice along this fringe, plotted in Fig.~\ref{fig:leak_floquet}(d), is then fitted to extract the leakage rate $L_1$.

However, this two-level SU(2) model is an approximation, as the CZ gate also induces a small population transfer to the other gate state $\ket{020}$. To justify our use of the simplified model, we quantified the impact of this parasitic leakage. We simulated the relative error on the extracted leakage rate $L_1$ as a function of the leakage to $\ket{020}$ (denoted $L_1^{\ket{2}}$). As shown in Fig.~\ref{fig:leak_floquet}(e), for typical experimental values of $L_1^{\ket{2}}$, this error is below 3\%. This confirms that the two-level approximation is a valid and highly accurate tool for extracting the primary leakage rate to the spectator state, and importantly, it does not affect the identification of the optimal off-point, which only relies on finding a minimum.

\subsection{Leakage Rate and Gate Error Extraction from XEB}
\begin{figure}[tbh]
    \centering
    \includegraphics[width=5.7cm]{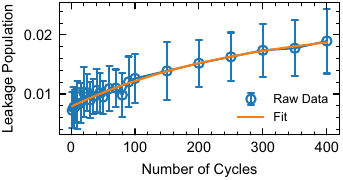}
    \caption{Leakage population dynamics of the spectator qubit $Q_{s_1}$ as a function of XEB cycles with the suppression protocol active. The data corresponds to the configuration where only $Q_{s_1}$ is tuned near resonance, extracted from the multi-spectator experiment presented in Fig.~4. The gate parameters are identical to those used in Fig.~3(d). Blue circles represent the experimental data, and the solid orange line indicates the fit to the rate equation.
}
    \label{fig:supp_leak_heat}
\end{figure}
To extract the leakage rate from the XEB data, we model the evolution of the population in the leakage subspace, $p_L(m)$, after $m$ cycles. The average population dynamics can be described by a phenomenological rate equation whose solution for a large number of cycles is:
\begin{equation}
    p_L(m) = \left(p_0 - p_\infty \right) \lambda_1^m + p_\infty,
    \label{eq:p_leak_fit}
\end{equation}
where $p_0 = p_L(0)$ is the initial thermal population, $\lambda_1 = 1-\left(L_1+L_2 \right)$, $L_1$ is the per-cycle leakage rate, $L_2$ is the per-cycle seepage rate (decay back to the computational subspace), and $p_\infty = \frac{L_1}{L_1+L_2}$ is the steady-state population. By fitting this model to the experimentally measured leakage population (after readout error correction), we can extract the physical rates $L_1$ and $L_2$.

The dynamical behavior of the leakage population is determined by the interplay between the initial thermal state $p_0$ and the steady-state limit $p_\infty$.
In our experiment, $p_0$ is fixed by the thermal equilibrium at the spectator's idle frequency.
The steady-state population $p_\infty$ is determined by both $L_1$ and $L_2$.
With the dynamic protocol applied, $L_1$ is suppressed to a low level ($\sim 10^{-4}$), making $p_\infty$ sensitive to the seepage rate $L_2$.
The seepage rate $L_2$ consists of two components: the energy relaxation time $T_1$ of the spectator qubit during the resonant interaction and the state-dependent leakage channel $|111\rangle \to |120\rangle$ associated with the excited spectator state $|1\rangle$ (discussed in Sec.~\ref{sec:supp_statedependent}).
Based on our simulations and the thermal population, the contribution from the $|111\rangle \to |120\rangle$ channel is estimated to be on the order of $10^{-6}$ as shown in Fig.~\ref{fig:supp_leakage}(b).
This is negligible compared to the contribution from $T_1$ relaxation, implying that $L_2$ is primarily governed by the effective $T_1$ during the XEB sequence.

Consequently, temporal fluctuations in $T_1$ can shift the relative magnitude of $p_\infty$ and $p_0$, resulting in different curve profiles.
In the dataset shown in Fig.~3(d) of the main text, the extracted $L_2$ corresponds to an $T_1 \approx 9.1~\mu\text{s}$ during the resonant interaction.
This fluctuation lowers $p_\infty$ below $p_0$, leading to a decaying curve.
Conversely, in other datasets with identical gate parameters, such as the one shown in Fig.~\ref{fig:supp_leak_heat}, a different effective $T_1$ results in $p_\infty > p_0$, yielding a monotonically increasing curve.
Both behaviors are fully consistent with the phenomenological rate model described by Eq.~\eqref{eq:p_leak_fit}. In addition, the leakage rate extracted from the dataset in Fig.~\ref{fig:supp_leak_heat} is $6(2) \times 10^{-5}$, which is consistent with the value reported in the main text. This confirms that the determination of the leakage rate $L_1$ remains robust against fluctuations in the seepage rate $L_2$.

The presence of leakage also affects the fidelity of operations within the computational subspace. The average fidelity of an XEB sequence of length $m$ can be modeled as:
\begin{equation}
   F(m) = A \lambda_1^m + B \lambda_2^m + C,
\end{equation}
where $\lambda_2$ is the depolarization parameter related to the coherent gate error. The average fidelity of a gate in the presence of leakage is given by~\cite{2018_leakage_cali}:
\begin{equation}
   \bar{F} = \frac{1}{d} \left[ (d-1)\lambda_2 + (1 - L_1) \right],
\end{equation}
where $d=4$ is the dimension of the two-qubit computational subspace. This formula allows us to understand how the leakage rate $L_1$ directly contributes to the average gate error measured by XEB. 

\subsection{Gate Error Analysis and Budget}
\begin{figure}[tbh]
    \centering
    \includegraphics[width=10.1cm]{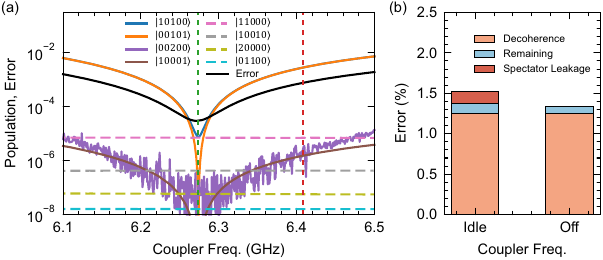}
    \caption{Detailed breakdown of leakage channels and gate error budget.
        (a) Simulated population dynamics in the two-excitation subspace.
        The system is initialized in the state $|10100\rangle \left(lc_ghc_ss \right)$.
        The blue solid line (labeled $|10100\rangle$) represents the aggregated total leakage population into the non-computational subspace.
        Solid lines indicate leakage channels sensitive to the spectator coupler frequency, while dashed lines represent frequency-independent background leakage.
        The green and red vertical dashed lines mark the optimal ``off'' frequency and the ``idle'' frequency, respectively.
        (b) Comparative error budget for the CZ gate.
        The total error is decomposed into spectator leakage, remaining coherent errors, and decoherence.
}
    \label{fig:supp_error_budget}
\end{figure}
The average gate error $\epsilon = 1-\bar{F}$ is directly impacted by leakage. The incoherent error from population loss contributes a term of $\frac{1}{4}L_1$ to the total error. Additionally, the leakage process induces a coherent error whose magnitude is also approximately $L_1$~\cite{barendsDiabaticGatesFrequencyTunable2019}. Summing these effects, the total gate error component arising from leakage, denoted $\epsilon_L$, can be approximated as:
\begin{align}
    \epsilon_L \approx L_1 + \frac{1}{4}L_1 = \frac{5}{4}L_1.
\end{align}
This model predicts that a reduction in leakage, $\Delta L_1$, should cause a corresponding reduction in the total gate error of $\Delta \epsilon \approx \frac{5}{4}\Delta L_1$.

We test this prediction with our experimental data. Applying our dynamic protocol reduced the measured leakage rate by $\Delta L_1 = 1.17\left(6 \right)\times10^{-3}$. The predicted reduction in gate error is therefore $\Delta \epsilon_{\text{pred}} = 1.46 \times 10^{-3}$. Experimentally, we measured a total gate error reduction of $\Delta \epsilon_{\text{exp}} = 1.8\left(10 \right)\times10^{-3}$. The predicted error reduction is consistent with the experimentally measured value within the measurement uncertainty. This agreement confirms that the observed fidelity improvement is quantitatively explained by the suppression of leakage and its associated coherent errors.

To systematically investigate the gate error budget and assess potential trade-offs introduced by the dynamic protocol, we performed a detailed error budget analysis using Schrodinger equation simulations, considering the full multi-level system.

As shown in Fig.~\ref{fig:supp_error_budget}(a), we analyzed the residual populations of non-computational states and the total gate error as a function of the spectator coupler frequency. The results indicate that the total leakage is dominated by the transition to the spectator state $|011\rangle$. Crucially, the residual population in the auxiliary gate state $|020\rangle$, which corresponds to a coherent error in the gate mechanism, exhibits a minimum synchronized with the spectator leakage at the optimal ``off'' frequency. 
Other background leakage channels, such as those to coupler modes, remain negligible and are insensitive to the control parameter.
Furthermore, the total gate error (black solid line) follows the same suppression profile as the two-excitation leakage.
This synchronization indicates that our protocol effectively restores the closed two-level dynamics between the computational state $|110\rangle$ and the hybridized gate state without introducing significant penalties in other error channels.  

Finally, we quantify the decoherent error contribution using the Lindblad master equation: 
\begin{align}
    \dot{\rho} = -i[H, \rho] + \sum_{k} \gamma_k \mathcal{D}[L_k]\rho,
\end{align}
where $L_k$ represents the collapse operators for relaxation and dephasing. By substituting the experimentally measured coherence times (Table.~\ref{Table: device params}), we derived the error budget shown in Fig.~\ref{fig:supp_error_budget}(b). The incoherent error sets a baseline of $1.25\%$. In the ``idle'' configuration, spectator leakage contributes a significant additional error. However, in the ``off'' configuration, this leakage contribution is eliminated, and the total error converges to the incoherent limit. These analyses demonstrate that the dynamic suppression scheme selectively eliminates the targeted spectator leakage without introducing significant side effects.

\subsection{Analysis of the Leakage Measurement Limit}

Our experimental results demonstrate leakage suppression to the order of $10^{-5}$. A natural question is what limits the ultimate performance and prevents reaching even lower values, such as the theoretical possibility of $10^{-8}$. The primary limitation is not the suppression method itself, but rather the resolution of our leakage measurement, which is fundamentally constrained by readout infidelity and qubit decoherence.

The steady-state leakage population, $p_\infty$, is determined by the balance between leakage and seepage: $p_\infty = \frac{L_1}{L_1 + L_2}$. To extract a small $L_1$, we need to accurately measure a small $p_\infty$. However, even after applying standard error mitigation techniques~\cite{2021_ibu}, residual readout errors impose a noise floor on population measurements. For our system, it is challenging to reliably measure population changes below the $10^{-2}$ level.

We can estimate the smallest leakage rate $L_1$ we can resolve by first calculating the seepage rate $L_2$. The seepage is dominated by the spectator qubit's energy relaxation during one XEB cycle, which consists of a single-qubit gate layer and a two-qubit CZ gate layer. The total seepage rate per cycle is the sum of contributions from each part:
\begin{equation}
    L_2 = \frac{t_{\text{1q}}}{T_1^{\text{idle}}} + \frac{t_{\text{cz}}}{T_1^{\text{res}}}.
\end{equation}
In our experiment, the single-qubit gate duration is $t_{\text{1q}} = 30~\text{ns}$ and the CZ gate duration is $t_{\text{cz}} = 40~\text{ns}$. The spectator qubit's relaxation time is $T_1^{\text{idle}} = 14.8~\mu\text{s}$ at its idle frequency and $T_1^{\text{res}} = 15.2~\mu\text{s}$ when tuned near resonance. These values yield a total seepage rate of $L_2 \approx 4.7 \times 10^{-3}$. If our measurement precision for the steady-state population is limited to a floor of $p_{\infty, \text{min}} \approx 0.01$, the minimum resolvable leakage rate $L_{1, \text{min}}$ can be estimated as:
\begin{equation}
    L_{1, \text{min}} = p_{\infty, \text{min}} \times (L_1 + L_2) \approx p_{\infty, \text{min}} \times L_2.
\end{equation}
This gives a resolution limit of $L_{1, \text{min}} \approx 4.7 \times 10^{-5}$. This estimated limit agrees well with our lowest measured leakage rate of $4 \left(3 \right) \times 10^{-5}$. This analysis strongly suggests that our dynamic suppression protocol is performing near the measurement limit of our current experimental setup, and the measured value represents an upper bound on the true suppressed leakage rate. 

This measurement-limited perspective also explains the results for multiple spectators. A simultaneous multi-qubit readout, required for the three-spectator case, suffers from lower fidelity due to factors like correlated errors. This degraded readout precision raises the measurement floor for the total leakage population, explaining why the suppressed total leakage in the three-spectator experiment is higher than the single-spectator limit. Therefore, achieving lower measured leakage rates, for both single and multiple spectators, would require significant improvements in readout fidelity and qubit coherence.

\subsection{Impact of Coupler-Coupler Coupling}
\begin{figure}[tbh]
    \centering
    \includegraphics[width=9.5cm]{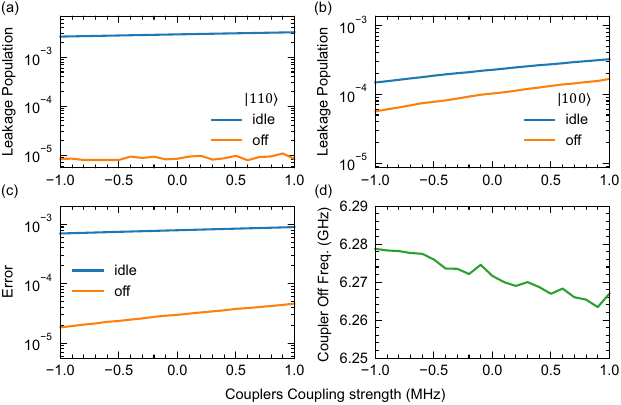}
    \caption{Impact of the parasitic coupling between the gate coupler and the spectator coupler on the leakage suppression protocol.
    (a), (b) Simulated leakage populations initialized from the two-excitation state $|110\rangle$ and the single-excitation state $|100\rangle$, respectively, as a function of the inter-coupler coupling strength.
    (c) Total gate error versus the inter-coupler coupling strength.
    The suppression performance remains robust across the swept range.
    (d) The shift in the optimal spectator coupler off frequency required to maintain leakage suppression as the coupling strength varies.
}
    \label{fig:supp_couplers_strength}
\end{figure}

The physical layout of the processor inevitably introduces a parasitic capacitive coupling between the gate coupler $C_g$ and the spectator coupler $C_s$. To evaluate the robustness of our protocol against this effect, we extend the theoretical model to include a direct interaction term $g_{cc}$. This coupling opens a third-order interaction pathway $Q_l \leftrightarrow C_g \leftrightarrow C_s \leftrightarrow Q_s$, which competes with the primary fourth-order pathway mediated by the qubit $Q_h$.

We derive the contribution of this new pathway using third-order perturbation theory. The effective coupling strength between the gate qubit $Q_l$ and the spectator qubit $Q_s$ is modified by an additive correction term:
\begin{equation}
    \tilde{g}_{ls} \approx g_{ls}^{(4)} + g_{ls}^{(3)},
\end{equation}
where $g_{ls}^{(4)}$ is the original fourth-order coupling derived in \eqref{eq:g1_derivation}. The third-order correction $g_{ls}^{(3)}$ arises from the direct coupler-mediated chain and is given by
\begin{equation}
\begin{aligned}
    g_{ls}^{(3)} = \frac{g_{lc_g} g_{c_gc_s} g_{sc_s}}{2} \left[ \frac{1}{\Delta_{sc_g}\Delta_{sc_s}} + \frac{1}{\Delta_{lc_g}\Delta_{lc_s}} \right].
\end{aligned}
\end{equation}
This result indicates that the presence of $g_{cc}$ effectively renormalizes the coupling parameters $\tilde{g}_{ls}$ but does not generate new types of interaction operators. Consequently, the block-diagonal structure of the effective Hamiltonian remains invariant. The condition for leakage suppression $g_{BD}=0$ is still mathematically solvable. The physical manifestation of this renormalization is simply a shift in the optimal spectator coupler frequency required to satisfy the condition. Therefore, the protocol is theoretically expected to be robust against such stray couplings.

We validated this robustness through master equation simulations by sweeping the inter-coupler coupling strength $g_{cc}$ across a realistic range. 
As shown in Fig.~\ref{fig:supp_couplers_strength}(a), the suppression of the primary leakage from the two-excitation state $|110\rangle$ remains highly effective. The residual leakage population at the optimal off point is consistently suppressed below $10^{-5}$, independent of the value of $g_{cc}$.
Fig.~\ref{fig:supp_couplers_strength}(d) confirms our theoretical analysis, showing that the optimal off frequency shifts linearly with $g_{cc}$ to compensate for the renormalized effective coupling.

Additionally, we examined the impact on the single-excitation subspace, where the third-order pathway could potentially enhance background leakage. Fig.~\ref{fig:supp_couplers_strength}(b) shows that while the leakage from $|100\rangle$ increases slightly with $g_{cc}$, it remains at the order of $10^{-4}$. This is negligible compared to the suppressed spectator leakage and other decoherence sources. 
Finally, Fig.~\ref{fig:supp_couplers_strength}(c) demonstrates that the total gate error remains stable across the swept range. These findings confirm that the direct coupling between couplers acts primarily as a parameter renormalization rather than a destructive mechanism, ensuring the practical applicability of our scheme in realistic circuit topologies.

\subsection{Independence and Additivity of Leakage from Multiple Spectator Qubits}
\begin{figure}[tbh]
    \centering
    \includegraphics[width=13.2cm]{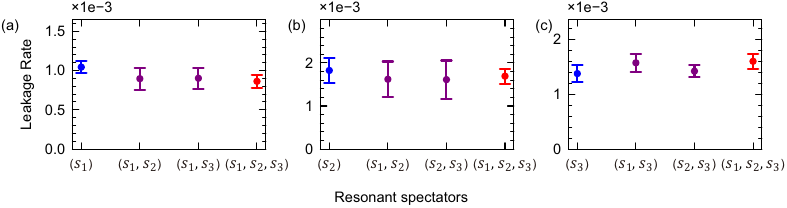}
    \caption{Measured leakage rates onto individual spectator qubits under different multi-spectator configurations. The x-axis denotes the set of spectators tuned to their near-resonant frequencies. (a) Leakage rate measured on $Q_{s_{1}}$. (b) Leakage rate measured on $Q_{s_{2}}$. (c) Leakage rate measured on $Q_{s_{3}}$. For each panel, only configurations where the target spectator is active are shown.
}
    \label{fig:supp_multi_spectator}
\end{figure}
In the main text, we investigated the total leakage from a CZ gate in the presence of up to three spectator qubits and demonstrated the scalability of our suppression technique. A key assumption underlying this scalability is that the leakage contribution from each spectator qubit is largely independent of the states or interactions of other spectator qubits. To validate this assumption, we performed a detailed analysis of the individual leakage pathways.

We systematically measured the leakage onto each spectator qubit across all eight possible configurations of the three spectators. A configuration is defined by which subset of the three spectators is tuned to its respective near-resonant frequency during the CZ gate, while the others remain at their idle frequencies. For each of the eight configurations, we ran a leakage amplification sequence and measured the final population in the excited states of $Q_{s_{1}}$, $Q_{s_{2}}$, and $Q_{s_{3}}$ individually.

The results are presented in Fig.~\ref{fig:supp_multi_spectator}. Fig.~\ref{fig:supp_multi_spectator}(a) shows the leakage population measured on $Q_{s_{1}}$ for all eight configurations. We observe that the leakage onto $Q_{s_{1}}$ remains nearly constant across all four configurations where $Q_{s_{1}}$ itself is tuned to resonance, regardless of whether $Q_{s_{2}}$ or $Q_{s_{3}}$ are also at their resonant frequencies. Similarly, when $Q_{s_{1}}$ is at its idle frequency, the leakage to it is negligible, independent of the other spectators' settings.

This behavior is consistently observed for the other two spectator qubits as well. Fig.~\ref{fig:supp_multi_spectator}(b) and (c) show the individual leakage measured on $Q_{s_{2}}$ and $Q_{s_{3}}$, respectively. In each case, the leakage to a specific spectator qubit is significant only when that qubit is tuned to its near-resonant frequency, and the magnitude of this leakage is almost identical across all configurations where this condition is met.

These findings strongly support the hypothesis that the leakage processes induced by each spectator qubit are independent to a very good approximation. The interaction between any two spectator qubits is a higher-order effect and can be safely neglected in our system. This validates the linear additivity of leakage rates used in the main text, where the total leakage is treated as the sum of individual contributions. This independence is a crucial property, as it ensures that leakage errors do not exhibit complex, correlated growth as the system size increases, simplifying both the characterization and mitigation of such errors in larger quantum processors.

\begin{figure}[H]
    \centering
    \includegraphics[width=13cm]{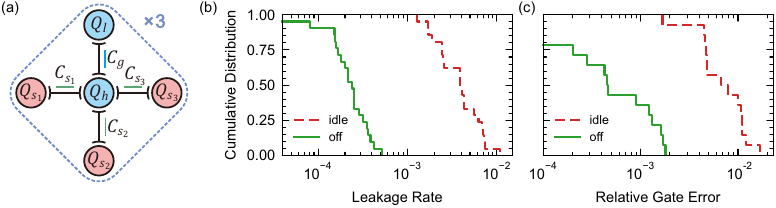}
    \caption{Statistical characterization of leakage suppression performance across multiple qubit regions.
    (a) Schematic of the 5-qubit experimental unit.
    We repeated the measurements on three distinct instances of this unit selected from the 72-qubit processor.
    (b) Cumulative distribution of the measured leakage rates for 21 different spectator configurations.
    The red dashed line represents the data with the spectator coupler at the idle frequency, while the green solid line represents the data at the optimized off frequency.
    (c) Cumulative distribution of the relative gate error.
    This metric quantifies the increase in gate error relative to the isolated baseline performance.
}
    \label{fig:statistics}
\end{figure}

\subsection{Statistical Analysis of Suppression Performance}

To evaluate the robustness and reproducibility of our protocol, we extended the experimental validation beyond the single 5-qubit region discussed in the main text.
We selected two additional 5-qubit regions on a new 72-qubit processor and performed the same characterization.
This brings the total dataset to three distinct physical regions.
Due to variations in fabrication, the new processor exhibits smaller maximum coupling strengths.
Consequently, we adjusted the CZ gate durations to $70$~ns and $80$~ns for these new regions, compared to $40$~ns in the main text.
These longer gate durations naturally lead to higher accumulated leakage when the spectator coupler is at the idle point, providing a more rigorous test for our suppression scheme.

We collected data for all 7 possible spectator configurations in each of the 3 regions, yielding a total of 21 statistical datasets.
Figure~\ref{fig:statistics}(b) displays the cumulative distribution of the measured leakage rates.
With the spectator couplers at the idle point, the average leakage rate is $4.4 \times 10^{-3}$.
Upon applying the dynamic suppression protocol, the average leakage rate is reduced to $2.5 \times 10^{-4}$.
This represents a mean suppression ratio of 17.6, confirming that the technique remains highly effective across different devices and parameter regimes.

Furthermore, we assessed the impact of leakage suppression on the overall gate fidelity.
Figure~\ref{fig:statistics}(c) shows the cumulative distribution of the relative gate error.
We define the relative gate error as the fractional increase in the total gate error compared to the baseline performance measured in the dispersive regime.
When the leakage is unmitigated, the gate error increases significantly.
In contrast, with the dynamic suppression active, the relative gate error remains low.
This statistical evidence confirms that our protocol consistently eliminates spectator-induced leakage and preserves the intrinsic gate fidelity across the processor.